\documentclass[%
 reprint,
 amsmath,
 amssymb,
 aps,
 prb,
 rmp,
 floatfix,
]{revtex4-1} %

\usepackage{amsmath}
\usepackage{amssymb}
\usepackage{graphicx}
\usepackage{subfigure}
\usepackage{dcolumn}
\usepackage{bm}
\usepackage{xcolor}

\begin{document}

\newcolumntype{L}[1]{>{\raggedright\arraybackslash}p{#1}}
\newcolumntype{C}[1]{>{\centering\arraybackslash}p{#1}}
\newcolumntype{R}[1]{>{\raggedleft\arraybackslash}p{#1}}

\preprint{APS/123-QED}

\title{A generalized variational principle with applications to excited state mean field theory}

\author{Jacqueline A. R. Shea$^1$}

\author{Elise Gwin$^{1}$}

\author{Eric Neuscamman$^{1,2,}$}%
\email{eneuscamman@berkeley.edu}

\affiliation{
${}^1$Department of Chemistry, University of California, Berkeley, CA, 94720, USA \\
${}^2$Chemical Sciences Division, Lawrence Berkeley National Laboratory, Berkeley, CA, 94720, USA
}

\date{\today}
\begin{abstract}
We present a generalization of the variational principle that is compatible with
any Hamiltonian eigenstate that can be specified uniquely by a list of properties.
This variational principle appears to be compatible with a wide range of electronic
structure methods, including mean-field theory, density functional theory,
multi-reference theory, and quantum Monte Carlo.
Like the standard variational principle, this generalized variational principle
amounts to the optimization of a nonlinear function that, in the limit of an arbitrarily flexible
wave function, has the desired Hamiltonian eigenstate as its global minimum.
Unlike the standard variational principle, it can target excited states and select individual
states in cases of degeneracy or near-degeneracy.
As an initial demonstration of how this approach can be useful in practice, we employ it to improve
the optimization efficiency of excited state mean field theory by an order of magnitude.
With this improved optimization, we are able to demonstrate that the accuracy of the corresponding
second-order perturbation theory rivals that of singles-and-doubles equation-of-motion
coupled cluster in a substantially broader set of molecules than could be explored by
our previous optimization methodology.
\end{abstract}

\maketitle

\section{Introduction}
\label{sec:intro}

While the ground state variational principle has acted as the cornerstone of electronic
structure theory for decades, its usefulness is limited by its focus on the lowest Hamiltonian eigenstates.
Certainly this reality has not prevented the development of powerful excited state
methods based on other principles, such as linear response methods, or even methods
based on the variational principle itself, such as state-averaging methods.
However, these methods rely on making additional approximations beyond those required
for the ground state theories from which they are derived.
Linear response of course assumes that the excited states are in some sense close to
the ground state in state space (specifically, it assumes that they live in the ground
state's tangent space),\cite{HeadGordon:2005:tddft_cis} 
whereas state-averaging assumes that important wave function
traits such as the molecular orbitals are shared by all states.\cite{state_averaging:werner_meyer:1981,state_averaging:sergio:2019,lan:2019:sscasscf}
In a huge variety of applications, these approaches have been successful.
However, there remain important areas --- such as charge transfer, core excitation, and
doubly excited states --- where these additional layers of approximation continue to
impair predictive power and where it would be desirable to construct excited state methods
that do not require them.\cite{krylov:eomccsd,piecuch:delta_eom_ccsd,Neuscamman:2016:var,Chris:2016:eom_hsvmc}

One route to doing so is to work with excited state variational principles, which can
fully tailor the flexibility of an approximate wave function ansatz to the needs of
an individual excited state.
Typically based on functional forms that involve squaring the Hamiltonian operator,\cite{Messmer:1970,umrigar1988optimized,Chris:2016:omega,VanVoorhis:2017:sigma_scf}
these approaches must either accept a higher computational scaling than
their ground state counterparts\cite{VanVoorhis:2017:sigma_scf}
or resort to statistical evaluation\cite{umrigar1988optimized,Chris:2016:omega}
or approximations to their functional forms.\cite{Jacki:2018:esmf}
These challenges in mind, it would be interesting if a class of exact excited state
variational principles could be formulated without the need to square this difficult operator.
In this paper, we present one such class, discuss its prospects for wide utility, and
show that it can be used to improve the efficiency of excited state mean field (ESMF) theory.\cite{Jacki:2018:esmf}

One seemingly inescapable difficulty with excited states and degenerate states is that
they are harder to specify uniquely than non-degenerate ground states.
Indeed, the latter can simply be specified by demanding the state of lowest energy,
a prescription that is both straightforward and widely applicable.
For excited states, defining the Hamiltonian eigenstate that one wants is much less
straightforward.
At the very least, one must say something more specific about it, such as where it is
in the state ordering or what its properties are like.
This specification may be relatively simple, such as specifying that one is interested
in the Hamiltonian eigenstate with energy closest to a given value, but clearly must
become more involved in cases with degeneracy or near-degeneracy.
Here, we will take the perspective that Hamiltonian eigenstates whose unique specification
requires making more precise statements about their properties be accommodated by crafting
a generalized variational principle in which these more precise specifications can be encoded.
For example, when dealing with degeneracy, uniquely specifying the desired stationary state
might be accomplished by specifying desired values for both the energy and dipole moment.
Even in cases that are not strictly degenerate, optimization may be easier if one can
make statements about properties other than the energy that help differentiate the state
from other energetically similar states.
Crucially, however, these statements should do no more than identify the state, and so we
will insist that the overall approach produce the same optimized wave function regardless of
the details of what properties were used to uniquely identify it.

Although we will argue below that this generalized variational principle (GVP) can be employed
in many areas of electronic structure and will point out parallels to recent work in
density functional theory\cite{chris:2019:vesdft}
and multi-reference theory,\cite{lan:2019:sscasscf}
we will in this study use ESMF theory
as an example in which the approach offers clear practical benefits.
In our previous study of ESMF, we coupled an approximate excited state variational
principle to a Lagrangian constraint that ensured the optimizations produce energy
stationary points.\cite{Jacki:2018:esmf}
While this approach allowed us to verify that ESMF theory could act as a powerful platform
for excited state correlation methods and helped to inspire the GVP that we introduce here,
it possessed a number of difficulties.
First, the method of Lagrange multipliers is, strictly speaking, a saddle-point method,\cite{lagrange_saddlepoints}
and so complications arise when coupling it to standard unconstrained quasi-Newton methods.
Second, we have found that, in practice, the approach suffers from poor numerical conditioning and can take
many thousands of iterations to converge, which offsets the advantages of Hartree-Fock cost scaling
with an unusually high prefactor.
Third, the original formulation was entirely based on the energy and
thus not appropriate for cases with degeneracy.
As we discuss below, all three of these difficulties can be addressed by
optimizing the ESMF wave function with a GVP.
The result is an order-of-magnitude speedup for ESMF
wave function optimization, moving the method firmly into the
regime where subsequent correlation methods, rather than the
mean field starting point itself, dominate the cost of making predictions.

The proposed GVP appears to
create a number of opportunities in other areas of electronic structure.
For example, many other approaches exist for optimizing the orbitals of weakly correlated
excited states, such as $\Delta$-SCF\cite{deltascf} and the more recently introduced $\sigma$-SCF.\cite{VanVoorhis:2017:sigma_scf,VanVoorhis:2019:hp_sigma_scf}
In the case of $\Delta$-SCF, the GVP's specification of desired properties is similar in spirit
to the maximum-overlap approach, and in practice may be able to compliment it by helping avoid variational collapse.
Using similar logic, it appears likely that the GVP will be immediately compatible with a very recent
excited-state-specific variant of CASSCF  that at present\cite{lan:2019:sscasscf} employs the
same type of Lagrange multiplier approach as in our original formulation of ESMF.
Likewise, so long as the properties used to specify the desired state can be statistically estimated
alongside the energy, it should be possible to realize a GVP approach within variational Monte Carlo.
Finally, as we will discuss briefly below, the proposed GVP can be used to define a constrained search
procedure similar to the ground state constrained search of Levy,\cite{Levy1979}
allowing formally exact density functionals to be defined for
excited states that can be specified uniquely by a list of properties.
This approach should allow a recent density functional extension of ESMF theory\cite{chris:2019:vesdft}
to be reformulated and placed on firmer theoretical foundations.

This paper is organized as follows.
First, we introduce the GVP, discuss why it should be applicable to
a wide range of wave function approximations, and show that it may
also have uses in density functional theory via a
constrained search  procedure.
We then review the ESMF wave function ansatz before discussing both our original
optimization approach and a more efficient approach based on a simple version of the GVP.
We then present results showing the advantages of the newer optimization approach,
use the improved optimizer to converge ESMF for a wider selection of
molecules than was previously possible, and analyze the accuracy of the
corresponding\cite{Jacki:2018:esmf} perturbation theory (ESMP2).
Finally, we provide a proof-of-principle example of how the use of properties other
than the energy can assist in optimization and in the face of energetic degeneracy.
We conclude with a summary of our findings
and some thoughts on future directions.

\section{Theory}
\label{sec:theory}

\subsection{A generalized variational principle}
\label{sec::gvp}

Although rigorous excited state variational principles
(i.e.\ functions whose global minimums are exact excited states)
can be constructed by squaring the Hamiltonian operator,\cite{Messmer:1970}
we will take a different approach here as working with
$\hat{H}^2$ is typically more difficult than working with
expressions involving only a single power of $\hat{H}$.
We begin by taking note of the fact that,
when the ansatz is chosen as the exact (within the orbital basis) full configuration
interaction (FCI) wave function, all of the Hamiltonian's eigenstates are
energy stationary points and all of the energy stationary points are Hamiltonian
eigenstates.
This reality is made plain by simply constructing the FCI energy gradient with
respect to the coefficient vector $\vec{c}$,
\begin{align}
\nabla E
 = \frac{\partial E}{\partial \vec{c}}
 = \frac{2}{|\vec{c}\hspace{0.7mm}|^2}\big( H - E \hspace{0.5mm}\big)\vec{c},
\end{align}
and recognizing that it is zero if and only if $\vec{c}$ satisfies the FCI
eigenvalue equation.
Thus, when attempting to construct a variational principle that yields an
exact excited state in the limit of an exact (FCI) ansatz, it is sufficient
to take an approach that searches only among energy stationary points and that,
in the exact limit, is guaranteed to produce the specific desired stationary point.
Of course, the investigator must know something about the state they want in order to
ask for it specifically, and so the approach will need a formal mechanism 
for defining which stationary point (i.e.\ energy eigenstate) is being sought.
Crucially, while working with energy stationary points does require differentiating
the energy, and minimizing a function that is itself defined in terms of the energy gradient
may require some second derivative information,
the necessary derivatives do not require squaring the Hamiltonian and can usually be
evaluated without increasing the cost scaling beyond what is already necessary for
evaluating the energy in the first place.\cite{Jacki:2018:esmf}
Recognizing these formal advantages, we thus seek to define a generalized variational
principle for ground, excited, and even degenerate states
in which the energy gradient is the central player and a very general mechanism is
provided for specifying which Hamiltonian eigenstate is being sought.

To begin, let us make an exact formal construction, after which we will discuss
how this construction may be converted into a practical tool.
First, take the wave function to be a linear combination
(with coefficient vector $\vec{c}\hspace{0.6mm}$) of
all the $N$-electron Slater determinants that can be formed
from the (finite) set of one-electron kinetic energy eigenstates that
(a) satisfy the particle-in-a-box boundary conditions 
for a large box whose edges are length $L$
and (b) have kinetic energy less than $Q L$,
where $Q$ is a fixed positive constant.
As the large box gets larger, this wave function ansatz
will eventually be able to describe any normalizable Hamiltonian
eigenstate to an arbitrarily high accuracy.
Next, choose a set of operators $\hat{B}_i$ and their
desired expectation values $b_i$ and define
a vector $\vec{d}$ of property deviations.
\begin{align}
  \vec{d} = \left\{ \hspace{1mm}
  \left< \hat{B}_1\right> - b_1, \hspace{3mm}
  \left< \hat{B}_2\right> - b_2, \hspace{3mm}
  \ldots \hspace{1mm} \right\}
\end{align}
Now, if this vector uniquely specifies an exact
Hamiltonian eigenstate, by which we mean that one
such eigenstate produces a lower norm for this vector
than any other Hamiltonian eigenstate, then that
eigenstate will be the result of the following
limit, which forms our generalized variational principle (GVP).
\begin{align}
\label{eqn:gvp}
  \lim_{L\rightarrow\infty} \hspace{1.2mm}
  \lim_{\mu\rightarrow0} \hspace{1.2mm}
  \min_{\vec{c}}
  \bigg(
  \mu \big| \vec{d} \hspace{0.7mm} \big|^2
  + (1-\mu) \big| \nabla E \big|^2
  \bigg)
\end{align}
Note the order of limits, in which we take the limit in $\mu>0$ for each value of $L$ as $L$ is made progressively larger. 
For any finite $L$, the properties of the system will be finite
regardless of the choice of $\vec{c}$, and so the largest possible norm for the
vector $\vec{d}$ will also be finite.
This implies that, as $\mu$ becomes small, the only states that stand
a chance of being the minimum are energy stationary states.
As the box gets bigger and bigger, the wave function approximation
becomes exact, and the energy stationary states tend towards the
exact Hamiltonian eigenstates.
By assumption, one of these eigenstates has a lower value for the
norm of $\vec{d}$ than the others, and so that eigenstate results
from the minimization, as desired.
While a non-degenerate ground state can be found
via a property vector containing only the energy
by setting the target energy to a very large negative number,
one can instead seek excited states by choosing other target
energies and furthermore can address degeneracies by adding
additional properties.

Although this formal definition works in principle, let us now
turn our attention to how it can be made useful in practice.
First, we replace the large-box FCI
wave function with an approximate wave function ansatz.
This removes the limit in $L$, which was in any case merely a way of
defining an explicit systematic approach towards an exact wave function.
In its place, we now have the idea of systematic improvability that usually
gets associated with variational principles:  as the approximate ansatz
becomes more and more flexible, we are guaranteed to recover the exact
eigenstate eventually.
Of course, as with the traditional variational principle, there will
be systems for which the approximate wave functions we can afford to
work with will not approach the exact limit closely enough to be useful.
For example, the Hartree-Fock Slater determinant is known to break symmetry
in unphysical ways in many situations upon energy minimization,\cite{coulsonfisher1949}
and we see no reason that similar qualitative failures should not occur for
too-approximate wave functions when using a GVP.
That said, the data provided below for ESMF theory suggest that there are
many cases where even relatively simple wave functions are flexible
enough to make the approach useful. 

In terms of affordability, notice that we have not relied on squaring the
Hamiltonian operator but have instead employed the norm of the energy
gradient with respect to the wave function's variational parameters.
If our wave function approximation allows for an affordable energy evaluation,
then automatic differentiation guarantees that the energy gradient,
as well as the gradient of the norm of the gradient
needed to perform the minimization,\cite{Jacki:2018:esmf}
can be evaluated at a constant multiple of the cost for the energy.
In the same way that we apply this approach to ESMF theory below, we
expect that a related recent approach\cite{lan:2019:sscasscf}
to excited-state-specific
CASSCF can also be reformulated
in terms of this GVP.
Of course, in practice, there may be faster ways of evaluating the
necessary gradients than resorting to automatic differentiation, but we
at least have that option in principle and so the worst-case scenario
for cost scaling should not be worse than the parent method.
As one additional comment on practical minimization, note that the
limit on $\mu$ will have to be discretized,
but as we discuss below in our application to ESMF theory, this does not
appear to create any significant difficulty.

At first glance, one might worry that, since the same eigenstate can be
specified by many different property deviation vectors, the approach may
give different results for different users who make different choices.
However, due to the limit on $\mu$, only energy stationary points will
result from the minimization.
So as long as the different property deviation vectors all specify the same
stationary point, they will produce the same results (under the usual assumptions
of nonlinear minimization methods not getting trapped in local minima).
Of course, there will be cases where it is not clear what to specify,
but in these cases it is not obvious that state-specific variational methods
should be used at all, for how does one design an optimization to find a
state about which nothing is known?
If something is known, and if other states that the desired state could get
confused with have been identified, then projections (even approximate ones)
against those states can be included in the property deviation vector
in an effort to uniquely specify the elusive state.
Of course, such an approach has its limits, and in systems with very dense
spectra, such as excitations inside a band of states in a solid,
state-specific methods are hard to recommend.

Before moving on to combining this general approach with ESMF theory, let us
make a short statement about density functional theory.
For a very large box (take a box as large as you need to make the following as
exact as you would like, e.g.\ set $L$ to one kilometer and $Q$ to one
megajoule per kilometer)
we define the density functional
\begin{align}
  G_{\mu}[\rho] = 
  \min_{\vec{c}\rightarrow\rho}
  \bigg(
  \mu \big| \vec{d} \hspace{0.7mm} \big|^2
  + (1-\mu) \big| \nabla E \big|^2
  \bigg)
\end{align}
in which we use Levy's constrained search\cite{Levy1979} approach of
searching over only those coefficient vectors $\vec{c}$ whose
wave functions have density $\rho$.
Minimizing this functional in the small-$\mu$ limit,
\begin{align}
  \lim_{\mu\rightarrow0} \hspace{1.2mm}
  \min_{\rho} \bigg( G_{\mu}[\rho] \bigg),
\end{align}
is then guaranteed to produce the exact
density for the Hamiltonian eigenstate specified by the supplied
property deviation vector.
Although minimizing $G$ by varying $\rho$ is an entirely impractical way of finding
the prescribed state's exact density, it is one way to do it,
and we therefore see that exact density functionals (meaning functionals of the density
that when minimized give the exact density)
exist for states that can be specified uniquely by property deviation vectors.
We make no attempt here to put precise bounds on how broad a class of states this is,
but we expect that it contains the vast majority of molecular excited states
that chemists have questions about.
Certainly, any non-degenerate molecular bound state falls in this category
as such states can be specified uniquely by their energy.

\subsection{The ansatz}
\label{sec::tsub_ansatz}

\begin{figure}
    \centering
    \includegraphics[scale=0.35]{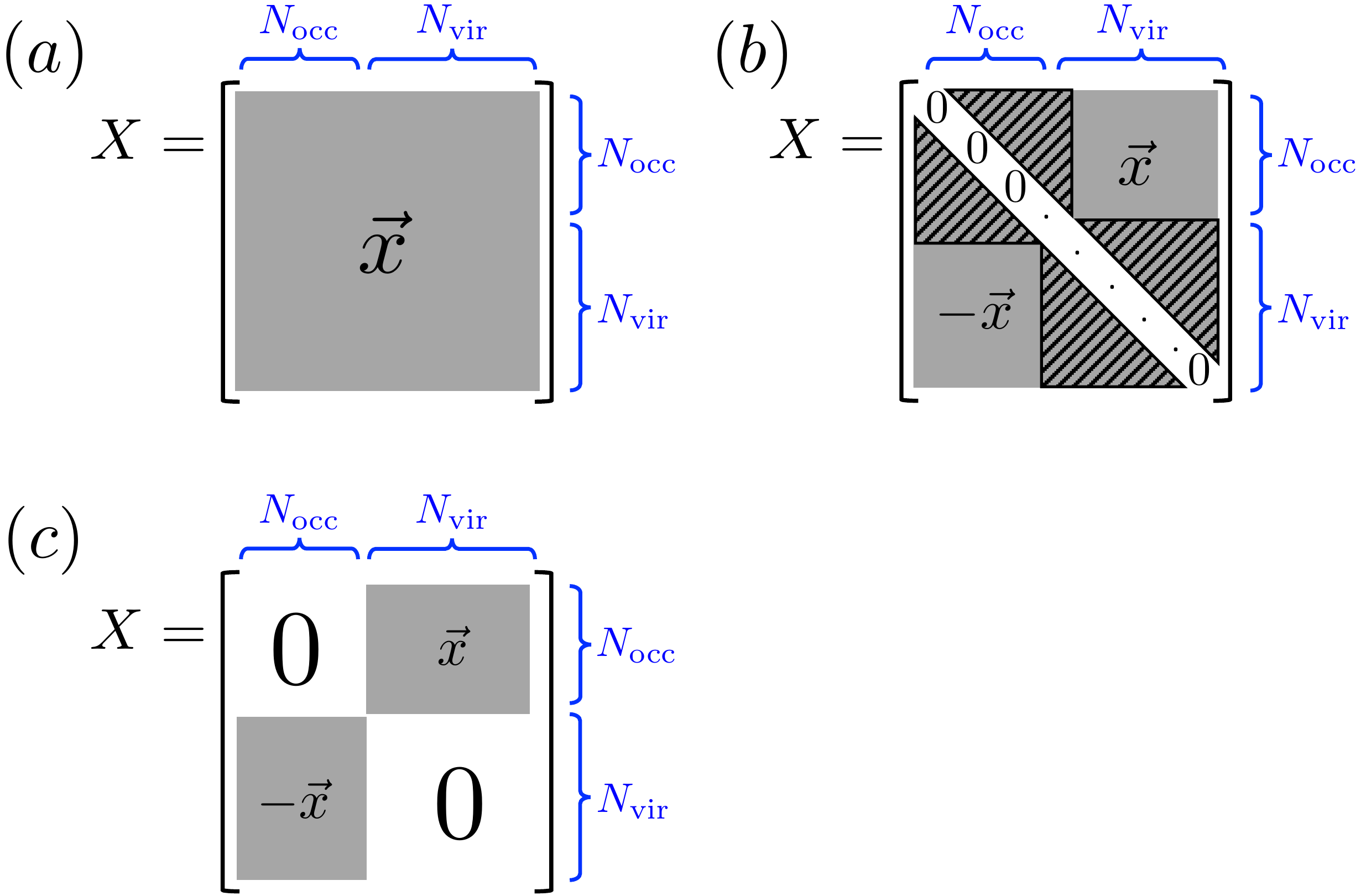}
    \caption{Gray areas represent non-zero variational parameters in the orbital rotation space. The black-striped triangles in (b) identify energy-invariant parameters. 
    The structure of the rotation coefficient matrix with no assumptions is shown in (a), using the property of anti-hermiticity in (b), and using the property of 
    invariance under occupied-occupied and virtual-virtual rotations in (c). }
    \label{fig:X_matrix}
\end{figure}

Turning now to the construction of a practical optimization
method for ESMF theory based on the above GVP, let us begin by
reminding the reader how ESMF theory defines its wave function
approximation.
\begin{align}
|\Psi\rangle = e^{\hat{X}} \Bigg( 
c_0|\Phi\rangle 
+ 
\sum_{ia} 
\sigma_{ia} 
\hat{a}_{a\uparrow}^\dagger \hat{a}_{i\uparrow}|\Phi\rangle 
+ 
\tau_{ia}
\hat{a}_{a\downarrow}^\dagger \hat{a}_{i\downarrow}|\Phi\rangle  \Bigg)
\label{eqn:wfn}
\end{align}
Here $|\Phi\rangle$ is the Restricted Hartree Fock (RHF) solution and is included to help maintain orthogonality between the excited state and the 
reference ground state, and the coefficients $\sigma_{ia}$ and $\tau_{ia}$ correspond to excitations of an up-spin and down-spin electron, 
respectively, from the $i$-th occupied orbital to the $a$-th virtual orbital. In a finite basis set of $N_{\text{bas}}$ spatial orbitals, the operator $\hat{X}$ is given by
\begin{align}
\hat{X} = \sum_{p<q}^{N_{\text{bas}}} X_{pq} \left( \hat{a}_{p}^\dagger  \hat{a}_{q} - \hat{a}_{q}^\dagger  \hat{a}_{p} \right),
\end{align}
in which $\hat{X}$ is real and restricted to be the same for up- and
down-spin spin-orbitals.
These choices ensure that the orbital relaxation operator given by $e^{\hat{X}}$ is unitary\cite{Helgaker:2000_unitary} and spin-restricted.
As only the elements above the diagonal of $\hat{X}$ matter here, the number of variational parameters is reduced from $N_{\text{bas}}^2$ as in 
Fig.\ \ref{fig:X_matrix}(a)
to $N_{\text{bas}}(N_{\text{bas}}-1)/2$ as in 
Fig.\ \ref{fig:X_matrix}(b). 
Furthermore, as noted by Van Voorhis and Head-Gordon, the energy of this type of singly excited wave function is 
invariant to rotations between occupied orbitals and to rotations between virtual orbitals.\cite{HF_Preconditioner:6:VanVoorhisGDM} 
Therefore, these energy-invariant rotation parameters lead to redundancy in the parameters such that two wave functions could have 
different values of $\vec{x}$, $c_0$, $\vec{\sigma}$, and $\vec{\tau}$, but have the same energy. As this would further complicate 
any numerical optimization strategy, we only allow for rotations between occupied and virtual orbitals, making $X$ an off-diagonal block matrix 
(see Fig.\ \ref{fig:X_matrix}(c)), 
reducing the number of variational rotation parameters to $N_{\text{occ}}N_{\text{vir}}$, and redefining $\hat{X}$
\begin{align}
\label{eqn:trunc_X}
\hat{X} = \sum_{i}^{N_{\text{occ}}} \sum_{a}^{N_{\text{vir}}}  X_{ia} \left(\hat{a}_{a}^\dagger  \hat{a}_{i} - \hat{a}_{i}^\dagger  \hat{a}_{a} \right).
\end{align}

Accounting for all of the variables in the ESMF wave function, for a system with a closed-shell ground state with $N_{\text{elec}}$ electrons in a 
basis of $N_{\text{bas}}$ orbitals, there are $N_{\text{occ}}=N_{\text{elec}}/2$ occupied orbitals, and  $N_{\text{vir}}=N_{\text{bas}}-N_{\text{occ}}$ 
virtual orbitals. 
The CIS-like coefficient vector $\vec{c}=\{ c_0, \vec{\sigma}, \vec{\tau} \}$ has $(1+2N_{\text{occ}} N_{\text{vir}})$ elements; 
the orbital rotation parameter vector $\vec{x}$ has $N_{\text{occ}} N_{\text{vir}}$ elements; thus, the ESMF wave function $\Psi(\vec{\nu})$, 
where $\vec{\nu}=\{\vec{c},\vec{x}\}$, has a total of $3N_{\text{occ}} N_{\text{vir}} +1$ variables.  

Despite being composed of multiple determinants, we assert that the ESMF wave function hews closely to the principle of a mean-field theory, 
as it retains only the minimum correlations necessary to describe an open shell excited state of a fermionic system. 
Like the mean field Hartree Fock wave function, the ESMF wave function captures Pauli exclusion via the antisymmetry of Slater determinants. 
However, more correlation is needed to describe an open shell excited state. Not only can the electrons in the open-shell arrangement 
not occupy the same {\it{spin}} orbital, these opposite-spin electrons cannot occupy the same {\it{spatial}} orbital, 
which is a strong correlation not present in a closed-shell ground state that we insist on capturing here, 
which immediately requires at least two Slater determinants. As including the entire set of single excitations keeps the 
approach general by ensuring that any such open-shell correlation can be captured and does not increase the cost scaling compared to 
using a single pair of determinants, we opt for the wave function above as our ansatz. 
We do note, however, that a two-determinant ansatz would likely be effective
in many cases, although making that simplification has the potential to
complicate the corresponding perturbation theory, a point we will return
to in discussing our results.
Either way, double and triple excitations --- where most of the weak correlation effects that would move us 
squarely away from the mean-field spirit are to be found --- are not included.

More accurate than CIS due to the added flexibility from the orbital rotation operator,\cite{HeadGordon:2005:tddft_cis}
yet less accurate than methods like EOM-CCSD that incorporate more electron correlation,\cite{krylov:eomccsd}
ESMF is best seen as a gateway towards quantitatively accurate descriptions of excited states rather than an accurate method in its own right. 
Like Hartree Fock theory,\cite{Barlett:2007:cc_rev,Helgaker:2000_HF_platform} its ultimate utility is intended as a 
platform upon which useful correlation methods can be built.
So far, three such methods have been investigated:
1) a DFT-inspired extension to ESMF whose preliminary testing\cite{chris:2019:vesdft} reveals a valence-excitation accuracy similar to 
that of TD-DFT but also the promise of significant advantages in charge transfer states,
2) an excited state analog of Moller-Plesset perturbation theory, ESMP2,\cite{Jacki:2018:esmf} which we show below to be highly competitive in accuracy with EOM-CCSD, 
and 
3) a state-specific complete active space self-consistent field (SS-CASSCF) approach whose orbital optimization mirrors that of ESMF and whose root-tracking 
approach is similar in spirit to the GVP discussed here.\cite{lan:2019:sscasscf}

While our original strategy for optimizing the ESMF ansatz achieved the
same $\mathcal{O}(N^4)$ scaling as ground state mean-field theory,
the actual cost of the optimization was unacceptably high.
In many cases, it was more expensive than working with our
fully-uncontracted version of ESMP2, whose cost scaling goes
as $\mathcal{O}(N^7)$.
In searching for more practical optimization methods, we have considered
the option of deriving Roothaan-like equations, but so far this approach
has not yielded a practical optimization strategy.
Instead, we have found two other approaches to be more effective, at least for now.
First, one can replace the BFGS\cite{BFGS:Broyden,BFGS:Fletcher,BFGS:Goldfarb,BFGS:Shanno}
minimization of our original Lagrangian
with an efficient Newton-Raphson (NR) approach, which handles
the strong couplings between ansatz variables and Lagrange multipliers more effectively.
Second, one can use the GVP introduced above to redefine the
optimization target function in a way that avoids Lagrange multipliers
entirely, at which point both BFGS and NR are considerably accelerated.
In the next section,
we will discuss the old and new target functions, after
which we turn our attention to comparing the relative merits of BFGS
and NR and how ESMF admits a useful finite-difference approach to the
latter.

\subsection{Target Functions}
\label{ssec::tsub_tf}

\subsubsection{Lagrange Multiplier Formalism}
\label{sssec::tsub_tf_lagrangian}

While Hartree-Fock theory uses Lagrange multipliers to
enforce orthonormality between the orbitals,\cite{Szabo-Ostland}
our original target function for optimizing the ESMF ansatz used
Lagrange multipliers to ensure that the optimization ended
on an energy stationary point even when we approximated
an $\hat{H}^2$-based variational principle to keep it affordable.
By minimizing
 \begin{align}
\label{eqn:esmf_lagrangian}
L_{\vec{\lambda}} = W + \vec{\lambda} \cdot \nabla E, 
\end{align}
in which $W$ is an approximated excited state variational principle
and $\vec{\lambda}$
are the Lagrange multipliers, we guarantee that the optimization preserves
the useful properties of energy stationary points, such as size-consistency
for a product-factorizable ansatz.
Note that, even if $W$ is not approximated, such properties can be violated
in the absence of the Lagrangian constraint.\cite{Jacki:2017:omega}
Note also that, since we rotate from already-orthonormal Hartree Fock orbitals, we do not need to add additional Lagrange multipliers for
orthonormality.
In practice, we approximated the excited state variational principle
\begin{align}
\label{eqn:exact_varprin}
W = \frac{\langle\Psi|(\omega - \hat{H})^2|\Psi\rangle}{\langle\Psi|\Psi\rangle}
\end{align}
whose global minimum is the exact (excited) energy eigenstate with energy closest
to $\omega$.\cite{Messmer:1969,Messmer:1970}
Somewhat surprisingly, we found that even the aggressive approximation 
\begin{align}
\label{eqn:approx_varprin}
W \approx (\omega - E )^2
\end{align}
was sufficient in practice.\cite{Jacki:2018:esmf} 

While this formalism allows us to produce a number of successful
optimizations in small molecules and achieves the desired cost
scaling, it does create multiple complications.
In particular, this Lagrangian is unbounded from below with respect to
the Lagrange multipliers, so simple descent methods like
standard BFGS cannot be applied directly.
Instead, we proceeded by minimizing the squared norm of the gradient of
$L_{\vec{\lambda}}$, i.e. $|\nabla L_{\vec{\lambda}}|^2$, and although this does not increase the cost
scaling, it leads to an additional layer of automatic differentiation
that increases the cost prefactor.
Worse yet, as we will make clear below, this strategy was very poorly
numerically conditioned, causing BFGS to require a large number of
iterations to converge.
In contrast, a NR algorithm can directly search for and locate saddle points of this Lagrangian target function, 
which are the solutions we actually seek, 
thus avoiding the need for an extra layer of derivatives.\cite{lagrange_saddlepoints}
Furthermore, NR helps with the speed of convergence due to its more
robust handling of second-derivative couplings.
However, whether working with NR or BFGS, we find it even more
effective to avoid Lagrange multipliers entirely by
reformulating the target function using the GVP.

\subsubsection{Generalized Variational Principle Approach}
\label{sssec::tsub_tf_dynamic_avg}

Consider instead an optimization target function that can be switched between the energy itself and a simple version of the GVP
in which the energy is the only property in the deviation vector.
\begin{align}
\label{eqn:Lmuchi}
L_{\mu\chi} = \chi\Big( \mu (\omega-E)^2 + (1-\mu) |\nabla E|^2 \Big)
              + (1-\chi) E
\end{align}
If we first consider the case where we set $\chi=1$,
we see that we have a simple version of Eq.\ (\ref{eqn:gvp})
in which the limit on $L$ has been replaced by using the approximate
ESMF ansatz.
This target function is bounded from below, and so a series of
optimizations in which $\mu$ is made progressively smaller and then
set to zero can immediately employ either BFGS or NR.
Once we are close to convergence, we can in the case of NR then
switch $\chi$ to zero and rely on NR's ability to hone in
on an energy saddle point.
Of course, in practice, it may be system specific when and if
switching $\chi$ to zero is advantageous.
If one does not switch $\chi$ to zero, then
it is important to realize that it is possible for an optimization
to end at a point where
$\nabla|\nabla E|^2 = 0$ but $|\nabla E|^2 \neq 0$,
so the user must be careful to verify at the end of the optimization
that the energy gradient is indeed zero as expected.
In the results below, we did not encounter any optimizations converging
to a stationary point of $|\nabla E|^2$ that was not a stationary point
of $E$, but such points clearly exist and so this simple check should be
standard procedure.

While the use of $(\omega-E)^2$ was seen in our previous approach as
an approximation to an excited state variational principle, the GVP
approach helps us see that this is not where the approximation lies.
Indeed, for non-degenerate states, this simple choice for the deviation
vector will give exact results when used with an exact ansatz.
From this perspective, it is clear that the approximation being made
is instead an ansatz approximation.
To be precise, the assumption is that the stationary
points of the ESMF ansatz are, for the states we seek, similar to
those of FCI, which is the same assumption that is made
when formulating the Roothaan equations to find an energy
stationary point of the Slater determinant in Hartree-Fock theory.
Thus, as in the ground state case, the central assumption is that
the relevant energy stationary point of the mean-field ansatz is
a good approximation for the exact Hamiltonian eigenstate.
Energy minimization (for the ground state) or the use of the GVP
(for any state) are simply means of arriving at the relevant
stationary point.

\begin{figure}
    \centering
    \includegraphics[scale=0.36]{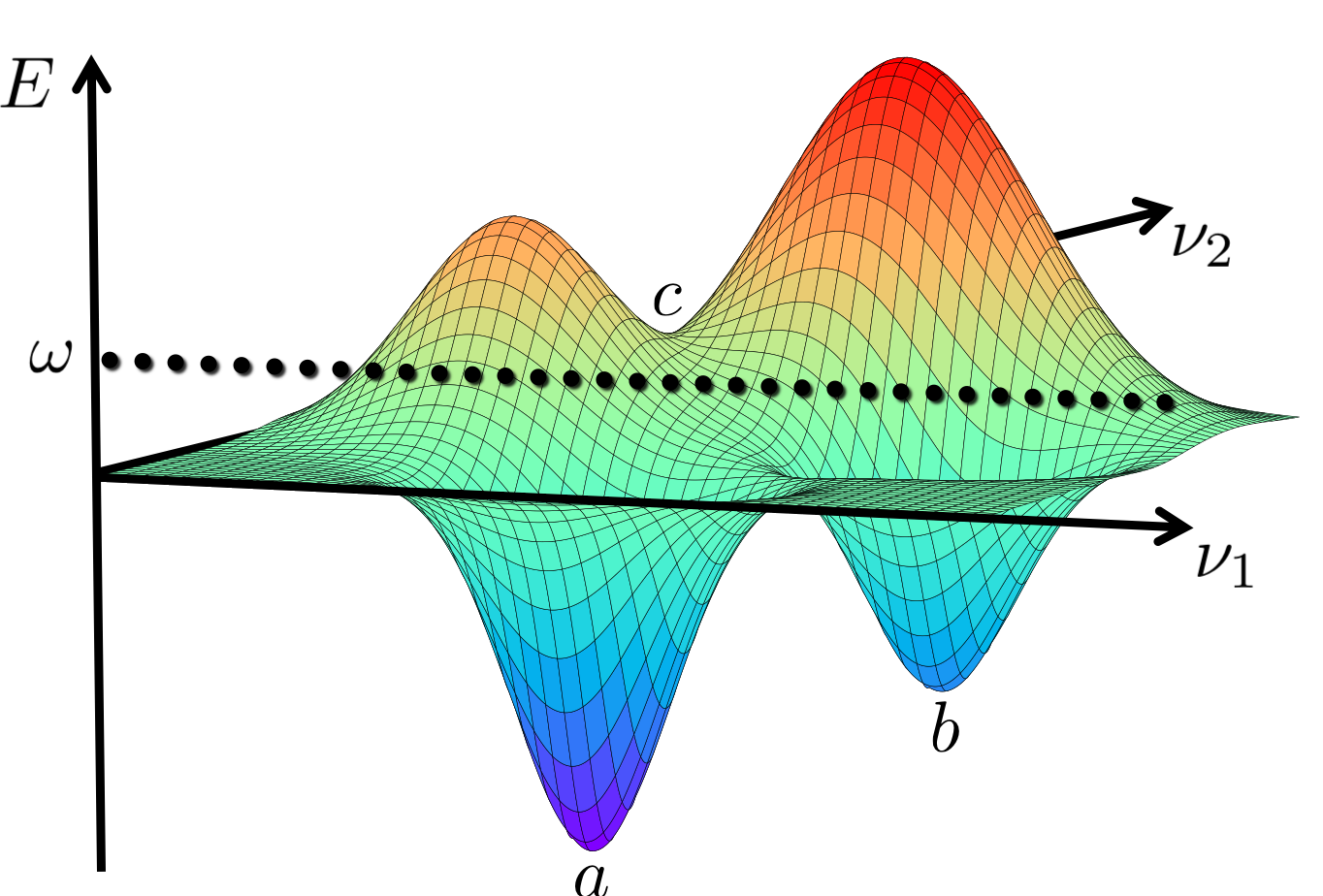}
    \caption{An idealized representation of the relationship between two variables, $\nu_1$ and $\nu_2$, in the ESMF wave function and the ESMF energy in a model system. 
    The global minimum at \textbf{a} is the lowest energy ground state of the system, while the two stationary points at \textbf{b} and \textbf{c} correspond to 
    excited states of the system, and can be individually resolved using the state-targeting parameter $\omega$ which is shown by the dotted line. }
    \label{fig:pes}
\end{figure}

To give a concrete idea of how this target function may be used in
practice, let us sketch out how we might use it to find
a particular stationary point (point $\bm{c}$) in 
Fig.\ \ref{fig:pes}.
Initially, we set $\chi=1$ and $\mu=\frac{1}{2}$
and select an $\omega$ value near to where we expect the
stationary point's energy to be.
We then take a series of NR steps seeking to minimize $L_{\mu\chi}$,
but in between each step we decrease $\mu$ by some small amount,
say $\frac{1}{10}$, until we reach $\mu=0$.
At this point, the GVP is assumed to have done its job
(locating the neighborhood of the correct stationary point)
and we switch $\chi$ to 0, thus converting the optimization
into a standard saddle point search (if this proves unstable
then $\chi$ can be held at 1 instead).
The potential advantage of this $\chi=0$ stage is that it
can employ well-known preconditioning methods
when solving the NR linear equation, such as the Hartree-Fock
energy hessian approximation
\begin{align}
\label{eqn:preconditioner}
    M_{ia,jb} = \frac{\delta_{ij}\delta_{ab}}{2(F_{aa}-F_{ii}+\Delta E)}
\end{align}
that is often used to accelerate ground state optimizations.\cite{HF_Preconditioner:1:Pople,HF_Preconditioner:2:Gordon,HF_Preconditioner:3:Backsay,HF_Preconditioner:4:Backsay,HF_Preconditioner:5:Almlof,HF_Preconditioner:6:VanVoorhisGDM} 
Here $F$ is the Fock matrix, and $\Delta E$ is the change
in energy between the last two optimization steps. 
In summary, what starts as a minimization method guided to the desired
stationary point by the GVP ends with a straightforward energy saddle
point search. 
Compared to the Lagrange multiplier approach, this approach has the
advantage that there are
no poorly-preconditioned Lagrange multipliers
and
the objective function is bounded from below. Thus, most standard
minimization methods can be used without the need for an additional
layer of derivatives, and the target function can convert into a form
that is easily preconditioned near the end of the optimization.

\subsection{Finite-Difference NR}
\label{ssec::tsub_om}

With the ability to evaluate gradients of both $L_{\vec{\lambda}}$
and $L_{\mu\chi}$ comes the option of employing a finite-difference
approximation\cite{Pearlmutter:1994:fastexactHv}
to the NR method (FDNR) that allows us to avoid the
expensive construction of Hessian matrices.
To begin, let us briefly review the standard NR method in order
to orient the reader and set notation.
In attempting to optimize $L$ with respect to the variables $\vec{\nu}$,
the NR approach approximates changes in $L$ to second order
in the yet-to-be-determined update to the current variable vector $\Delta\vec{\nu}$.
\begin{align}
    \label{eqn:nr_derivation_start}
    L(\vec{\nu} + \Delta \vec{\nu} )
    &\approx
      L(\vec{\nu})
    + \nabla L \cdot \Delta \vec{\nu} 
    + \frac{1}{2} \Delta \vec{\nu}^T \cdot \mathbf{H} \cdot \Delta \vec{\nu} 
\end{align}
Here, the Hessian is
$\mathbf{H}_{ij}\equiv \partial^2 L / \partial \nu_i \partial \nu_j $.
Setting the gradient of this approximation (w.r.t.\ $\Delta\vec{\nu}$)
to zero and solving for $\Delta\vec{\nu}$ gives an estimate for the
variable change that would lead to $L$ being stationary.
\begin{align}
    \label{eqn:nrupdate}
    \mathbf{H} \cdot \Delta \vec{\nu} \approx - \nabla L
\end{align}
If $\mathbf{H}$ can be explicitly constructed and inverted, then
one simply does so in order to solve for the update,
but in many situations (including ours) the explicit
construction of the Hessian is prohibitively expensive.
Instead, we follow Pearlmutter\cite{Pearlmutter:1994:fastexactHv}
and solve Eq.\ (\ref{eqn:nrupdate}) via a Krylov subspace method
(we use GMRES\cite{GMRES:Saad_Schultz})
in which the matrix-vector product is formed efficiently via
a finite-difference of gradients.
Noting that the key matrix-vector product is
\begin{align}
\label{eqn:hess_mat_vec}
[\hspace{0.8mm} \mathbf{H} \cdot \Delta\vec{\nu} \hspace{0.8mm} ]_i &=
\sum_j \frac{\partial^2 L}{\partial \nu_i \partial \nu_j } \Delta\nu_j
\end{align}
we compare to the differentiation of a first-order Taylor
expansion of $L$.
\begin{align}
L(\vec{\nu} + \Delta\vec{\nu}) &\approx L(\vec{\nu}) + \sum_j \frac{ \partial L(\vec{\nu}) }{\partial \nu_j } \Delta\nu_j \\
\label{eqn:dTEL}
\frac{ \partial L(\vec{\nu} + \Delta\vec{\nu}) }{\partial \nu_i } &\approx \frac{ \partial L(\vec{\nu}) }{\partial \nu_i } + \sum_j \frac{\partial^2 L(\vec{\nu})}{\partial \nu_i \partial \nu_j } \Delta\nu_j
\end{align}
Combining Eqs.\ (\ref{eqn:hess_mat_vec}) and (\ref{eqn:dTEL}),
one arrives at a simple approximation for the matrix-vector product
in the form of a gradient difference.
\begin{align}
\label{eqn:fda_i}
[ \hspace{0.8mm} \mathbf{H} \cdot \Delta\vec{\nu} \hspace{0.8mm} ]_i &\approx \frac{ \partial L(\vec{\nu} + \Delta\vec{\nu}) }{\partial \nu_i } - \frac{ \partial L(\vec{\nu}) }{\partial \nu_i } 
\end{align}
Note that, if $\Delta\vec{\nu}$ is not small enough to justify the
Taylor expansion, we can exploit the linearity of the matrix-vector
product to make the finite-difference more accurate by scaling the
vector down and then scaling the resulting matrix-vector product back up.
\begin{align}
\label{eqn:fda}
\mathbf{H} \cdot \Delta\vec{\nu}  &\approx \frac{1}{\epsilon} \Big( \nabla  L(\vec{\nu} + \epsilon \Delta\vec{\nu}) - \nabla L(\vec{\nu}) \Big) 
\end{align}
In a given FDNR iteration, $\nabla L(\vec{\nu})$ is evaluated once and
stored, so that each additional finite-difference estimate of a
matrix-vector product requires only a single additional gradient evaluation. 
Although in principle an even more accurate finite-difference can
be achieved at the cost of two gradients per matrix-vector multiply
via the symmetric finite-difference formula, we have not found this
to be advantageous.
As our results below demonstrate, the simpler one-gradient approach is
already a very accurate approximation.

\section{Computational Details}

Calculations in this work include timing comparisons between the previously discussed ESMF optimization methods and vertical excitation 
benchmarks of ESMF and ESMP2 against a range of single-reference excited state methods. 
Generally, generating initial guesses for an ESMF calculation is straightforward and does not necessitate any post-HF calculations. 
An initial RHF calculation computes the orthonormal HF orbitals which are used unrotated as the initial ESMF orbitals, i.e.\ $X=0$, 
and therefore the initial unitary transformation matrix is simply the identity matrix. Often, the dominant configuration state function (CSF) 
in an excitation is sufficient as the initial guess for $\vec{c}$. For example, using the notation from Eq.\ (\ref{eqn:wfn}), 
the initial guess for a singlet excitation from the $i$-th to the $a$-th orbital is simply 
$|\Psi\rangle = \frac{1}{\sqrt{2}} \mathbf{I} ( \hat{a}_{a\uparrow}^\dagger \hat{a}_ {i\uparrow} |\Phi\rangle +  \hat{a}_{a\downarrow}^\dagger \hat{a}_ {i\downarrow} |\Phi\rangle )$. 
Choosing $\omega$ is system dependent and requires some intuition, but even a rough approximation to the excitation energy from experiment, 
TD-DFT, or a CIS calculation is often sufficient because the energy stationary point criteria ensures that we will recover 
a size-consistent solution regardless of our choice of $\omega$. If a rough approximation of the excited state 
energy is not feasible, one can slowly increase $\omega$ over the course of several optimizations and identify 
an entire spectrum of excited states. In this specific survey, the former approximation process was used to generate the majority 
of initial ESMF guesses. For a few systems, however, the initial guesses were generated from the most important single excitation coefficients 
of the EOM-CCSD wave function, and $\omega$ was calculated using
\begin{align}
    \omega = E_{RHF} + \Delta E_{EOM-CCSD} - \frac{0.5}{27.211}
\end{align}
where $E_{RHF}$ is the RHF ground state energy and $\Delta E_{EOM-CCSD}$ is the EOM-CCSD vertical excitation energy. 
The energy correction is included since ESMF tends to underpredict excitation energies by approximately 0.5 eV 
relative to high level benchmarks. These few initial guesses built from EOM-CCSD results were only necessary to confirm 
that both methods were describing the same excitation and ensure fair comparisons and evaluations between them. 
The ground state references for ESMF and ESMP2 excitation energies are RHF and MP2, respectively.

Timing data reported in this work was produced on one 24-core node of the Berkeley Research Computing Savio cluster.
Note that this timing data is meant to compare between different optimization methods for ESMF that use the same Python-based Fock 
build code, and are not intended to represent production level timings.
Work is underway on a low-level implementation that exploits a faster Fock build, but this is not the focus of the current study. 
For FDNR-$L_{\mu\chi}$ timing data reported in Sec.\ \ref{ssec::rsub_tf}, $\mu$ was set to $\frac{1}{2}$ in the first iteration, 
$\frac{1}{4}$ in the second iteration, and $0$ in all subsequent iterations, and $\chi$ was switched from $1$ to $0$ after the first 10 iterations.   
All calculations were completed under the frozen core approximation and most in the cc-pVDZ basis set;\cite{6-31g,Dunning:1989:basis_for_corr} exceptions to the latter 
include the Hessian data reported in 
Fig.\ \ref{fig:hessian_heatmap} 
and the LiH system in Sec.\ \ref{ss:lih_property_vec} which used the minimal STO-3G basis set,\cite{sto-3g:1969,sto-3g:1976} and CIS(D) benchmark data which 
employed the rimp2-cc-pVDZ auxiliary basis set.\cite{rimp2-cc-pvdz} With MOLPRO version 2019.1,\cite{MOLPRO-WIREs,MOLPRO_brief} we optimized the geometries of a set of small 
organic molecules at the B3LYP/6-31G* level of theory.\cite{b3lyp:1,b3lyp:2,6-31Gstar:1988,6-31Gstar:1991}
Explicit geometry coordinates and the main CSFs contributing to each excitation are provided in the Supporting Information (SI). 
Moving to the cc-PVDZ basis set, we then performed ground state Restricted Hartree-Fock calculations to compute the initial orbitals used in ESMF, and CIS, 
MP2, and EOM-CCSD calculations for later benchmarking. CIS(D), TDDFT/B3LYP, and TDDFT/$\omega$B97X-V\cite{wB97X-V:initial}
excitation energies were computed with Q-Chem version 5.2.0\cite{QChem:2013}, and $\delta$-CR-EOM-CC(2,3)\cite{piecuch:delta_eom_ccsd} 
calculations were performed with GAMESS.\cite{gamess1} As some theories may lead to different orbital energies and thus orbital orderings, 
for each theory that used a different molecular orbital basis, we used Molden\cite{molden:2000,molden:2017} to plot and compare the main 
orbitals involved in each excitation to ensure we were comparing the same state between theories.

\section{Results}
\label{sec:results}

\begin{table}[t]
\caption{
For different molecules and stages in a NR optimization in the cc-pVDZ basis, we show the number of orbitals in the basis set $N_{\text{bas}}$, the time it takes to 
construct $\mathbf{H}_{\text{ex}}$ and then apply it to
$\nabla L_{\mu\chi}$ one hundred times, the time it takes to estimate the same one hundred matrix-vector products via Eq.\ (\ref{eqn:fda}),
and the average relative error associated
with Eq.\ (\ref{eqn:fda}).
}
\label{tb:fd_accuracy_timing}
  \begin{tabular}{llcccccc}
    \hline \hline	\rule{0pt}{3ex}
  &  &     &   NR        & $\mathbf{H}_{\text{ex}}$ time    & $\mathbf{H}_{fd}$ time&        relative        \\
  &Molecule  &$N_{\text{bas}}$& iter. & (min)              & (min)         & error  \\
  \hline \rule{0pt}{4ex}
 & Water        & 23 & 0  & 0.40 & 0.39  & $2.4 \times 10^{-6}$  \\ 
 &              &    & 5  & 0.40 & 0.39  & $6.0 \times 10^{-6}$ \\ 
 &              &    & 10 & 0.40 & 0.39  & $6.0 \times 10^{-6}$ \\ 
 \rule{0pt}{4ex}
 & Ammonia      & 28 & 0  & 0.97 & 0.79  & $2.4 \times 10^{-6}$   \\ 
 &              &    & 10 & 1.00 & 0.82  & $6.1 \times 10^{-6}$ \\ 
 &              &    & 30 & 1.03 & 0.93  & $6.1 \times 10^{-6}$ \\ 
 \rule{0pt}{4ex}
 & Formaldehyde & 36 & 0  & 4.52 & 2.08 &  $2.8 \times 10^{-6}$   \\ 
 &              &    & 10 & 4.53 & 2.08 &  $6.5 \times 10^{-6}$ \\ 
 &              &    & 18 & 4.53 & 2.08 &  $6.5 \times 10^{-6}$ \\ 
 \rule{0pt}{4ex}
 & Methanimine  & 41 & 0  & 9.02 & 3.62 &  $2.6 \times 10^{-6}$   \\ 
 &              &    & 10 & 9.22 & 3.81 &  $6.5 \times 10^{-6}$ \\ 
 &              &    & 30 & 9.38 & 3.77 &  $6.5 \times 10^{-6}$ \\ 
 \hline \hline
 \end{tabular}
\end{table}

\subsection{Assessing the Finite-Difference Approximation}
\label{ssec::rsub_fd}

While the numerical efficiency offered by the FDNR approach is welcome, its overall applicability depends on the accuracy of its finite-difference approximation.
To quantify its accuracy we computed the exact Hessian of $L_{\mu\chi}$, $\mathbf{H}_{\text{ex}}$, with Tensorflow's automatic differentiation software\cite{tensorflow}
for water, ammonia, formaldehyde, and methanimine, at the beginning, middle, and end of the optimization. 
We then constructed a ``finite-difference Hessian'' of  $L_{\mu\chi}$, $\mathbf{H}_{\text{fd}}$, by applying Eq.\ (\ref{eqn:fda}) to the columns of the identity matrix. 
In Table \ref{tb:fd_accuracy_timing}, we report
1) the number of orbitals in these systems with the cc-pVDZ basis set, 
2) the time in minutes required to build  $\mathbf{H}_{\text{ex}}$ once and compute the matrix-vector product $\mathbf{H}_{\text{ex}} \nabla L_{\mu\chi}$ one hundred times, 
3) the time in minutes required to compute $\mathbf{H}_{\text{fd}} \nabla L_{\mu\chi}$ using Eq.\ (\ref{eqn:fda}) one hundred times, 
and 4) the relative error between the results of applying either $\mathbf{H}_{\text{fd}}$ or $\mathbf{H}_{\text{ex}}$ to $L_{\mu\chi}$,
i.e.\ $ \big| \mathbf{H}_{\text{fd}}\nabla L - \mathbf{H}_{\text{ex}}\nabla L \big| / |\mathbf{H}_{\text{ex}}\nabla L|$. 
The data demonstrate that the finite-difference approach to applying the Hessian matrix has much more favorable scaling than building and applying the exact Hessian. In fact, the 
cost of building the exact Hessian scales so rapidly that, for
cyclopropene (59 orbitals in the cc-pVDZ basis),
$\mathbf{H}_{\text{ex}}$ could not be computed
in under two hours on a NERSC Cori Haswell node.
Additionally, the relative errors assure us that the finite-difference
approximation is highly accurate, and so we have used it in all
of the iterative FDNR optimizations discussed below.

\subsection{Comparing Optimization Strategies}
\label{ssec::rsub_tf}

\begin{figure}
    \centering
    \includegraphics[scale=0.43]{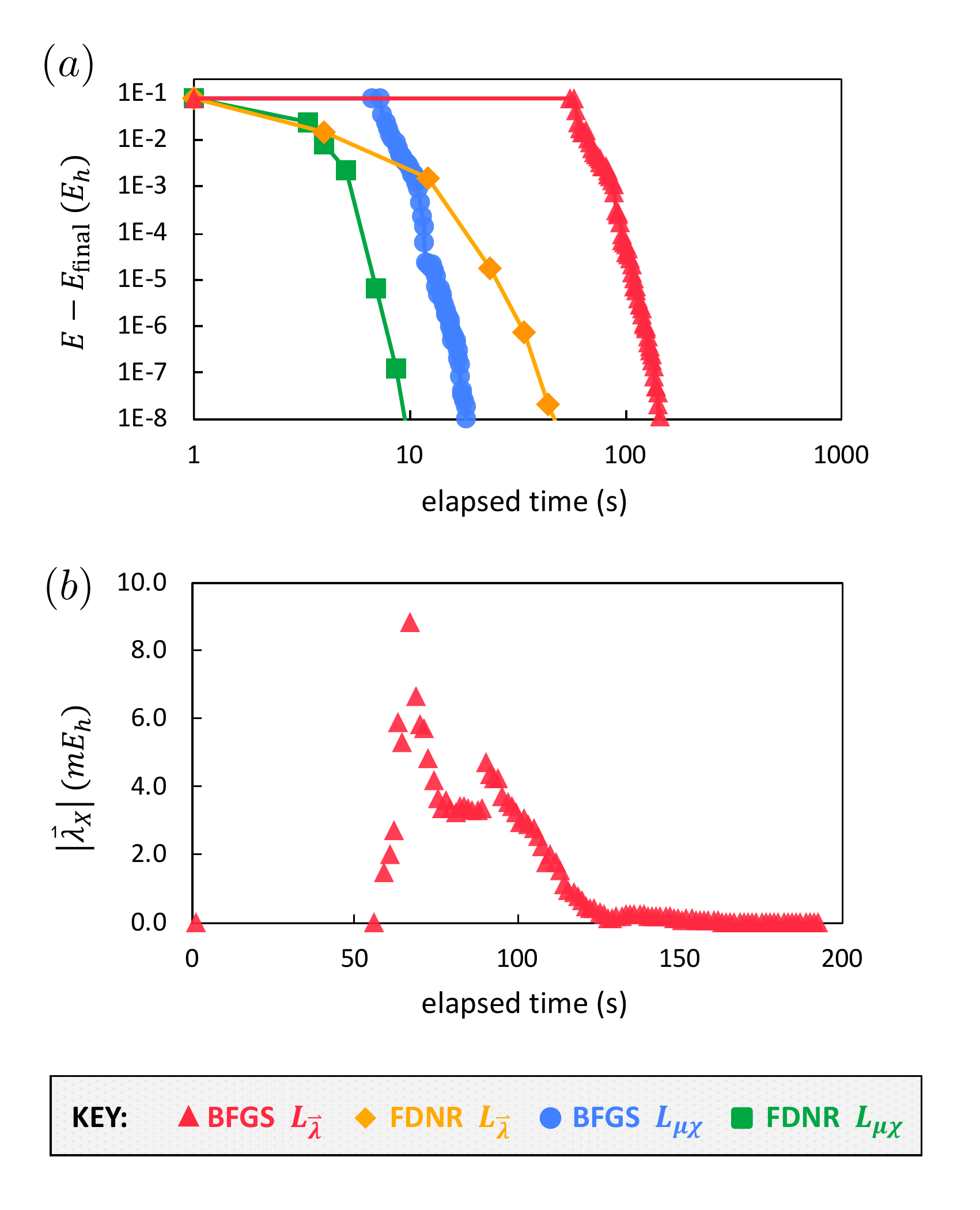}
    \caption{Two plots are shown for different optimizations of the HOMO-LUMO singlet excitation in cc-pVDZ 
    water.
    In (a), the difference between the current and converged excited state energy in Hartree is compared to the elapsed optimization time for the BFGS-$L_{\vec{\lambda}}$ (red triangles),  
    FDNR-$L_{\vec{\lambda}}$ (orange diamonds), BFGS-$L_{\mu\chi}$ (blue circles), and FDNR-$L_{\mu\chi}$ (green squares) optimization strategies.  In (b), the norm (in millihartrees) of the 
    Lagrange multipliers associated with the orbital rotation parameters is shown over the course of an BFGS-$L_{\vec{\lambda}}$ optimization. Note that the elapsed time is plotted on a 
    log scale in (a) but on a linear scale in (b) and that each marker represents one iteration in the associated algorithm. }  
    \label{fig:timings_water}
\end{figure}

\begin{figure}
    \centering
    \includegraphics[scale=0.43]{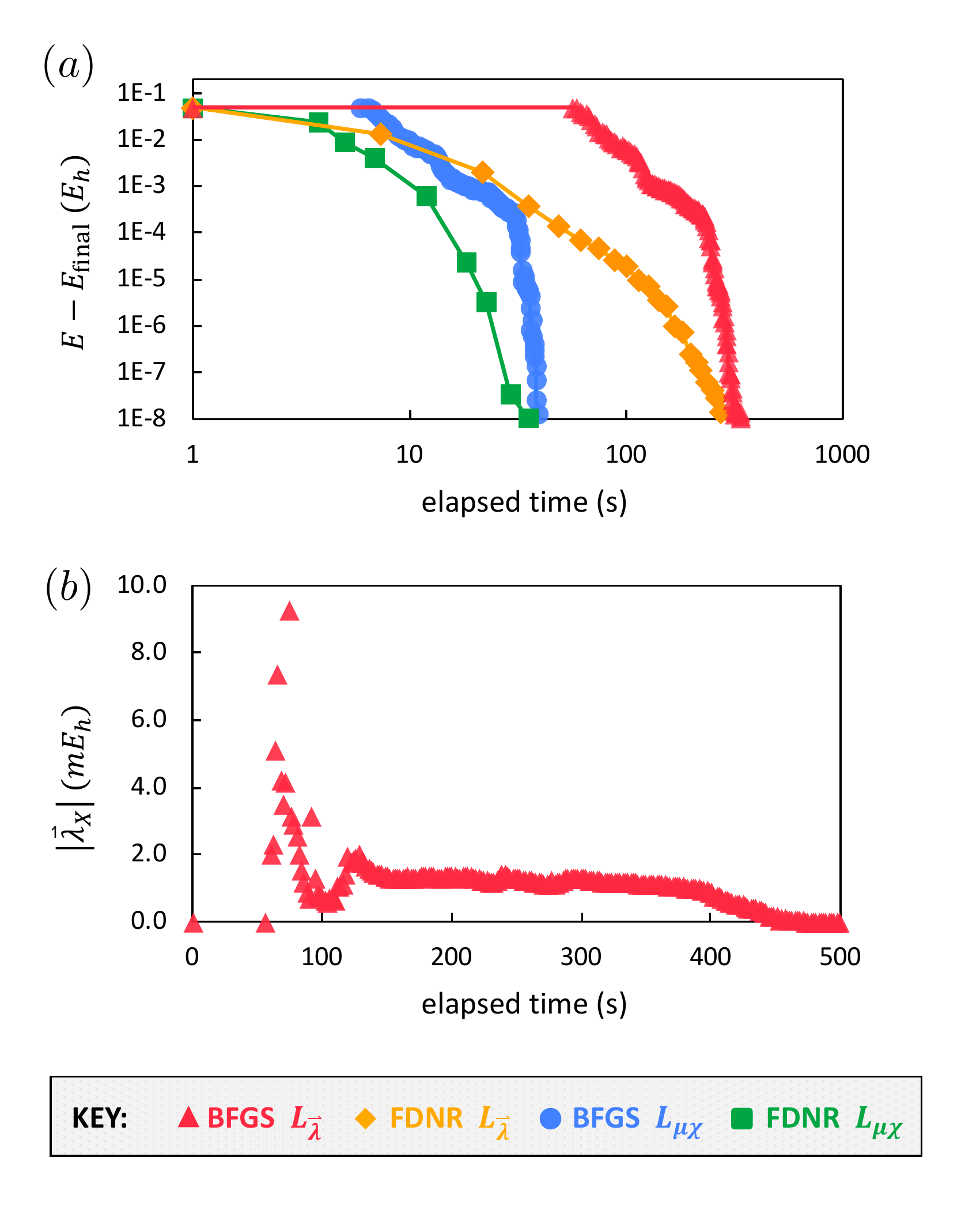}
    \caption{For the HOMO-LUMO singlet excitation in cc-pVDZ carbon monoxide, the difference between the current and converged excited state energy in Hartree is compared across optimization 
    strategies in (a) and the norm (in millihartrees) of the Lagrange multipliers associated with the orbital rotation parameters is shown over the course of an BFGS-$L_{\vec{\lambda}}$ 
    optimization in (b).}  
    \label{fig:timings_co}
\end{figure}

\begin{figure}
    \centering
    \includegraphics[scale=0.43]{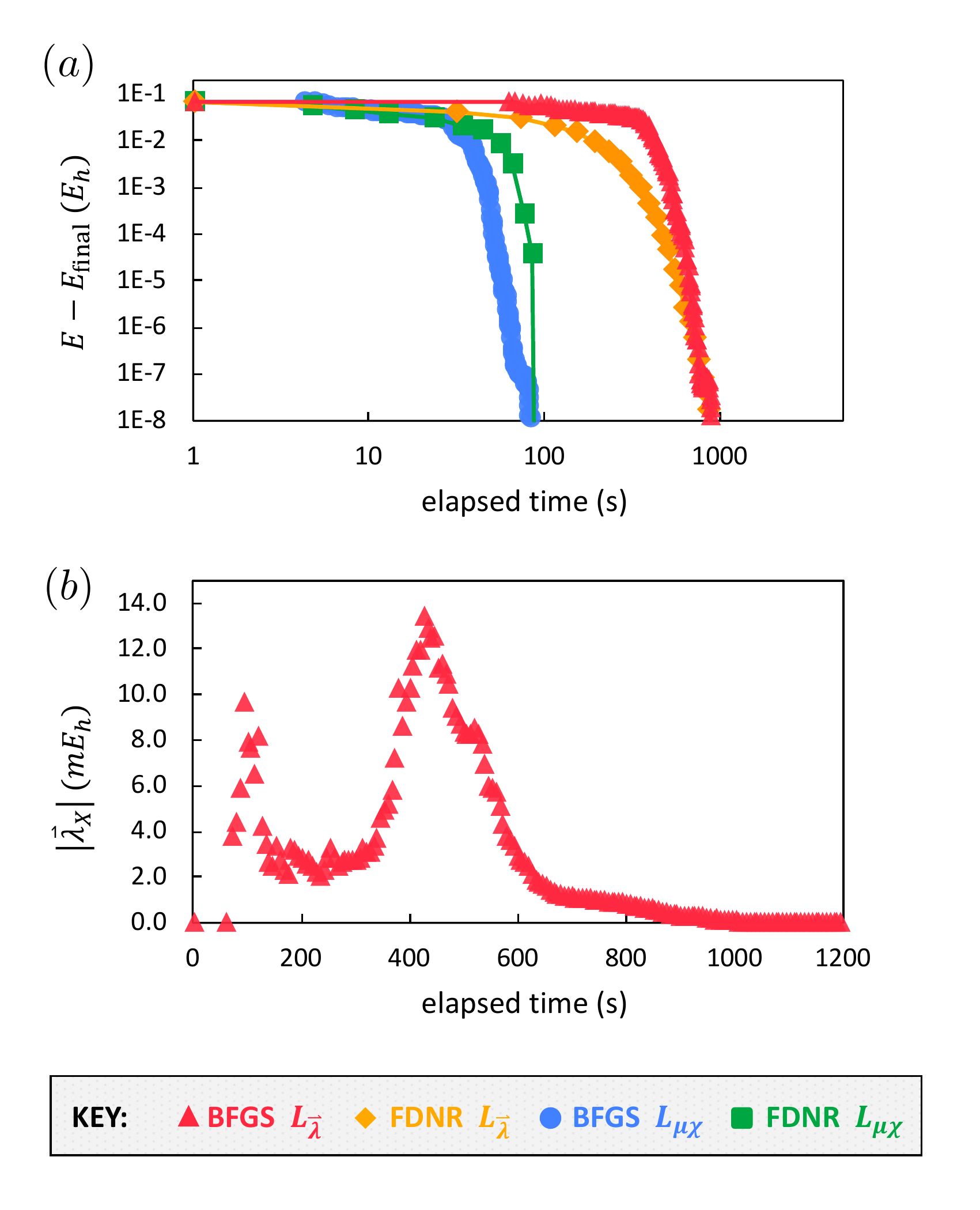}
    \caption{For the HOMO-LUMO singlet excitation in cc-pVDZ formaldehyde, the difference between the current and converged excited state energy in Hartree is compared across optimization 
    strategies in (a) and the norm (in millihartrees) of the Lagrange multipliers associated with the orbital rotation parameters is shown over the course of an BFGS-$L_{\vec{\lambda}}$ 
    optimization in (b).}  
    \label{fig:timings_formaldehyde}
\end{figure}

\begin{figure}
    \centering
    \includegraphics[scale=0.43]{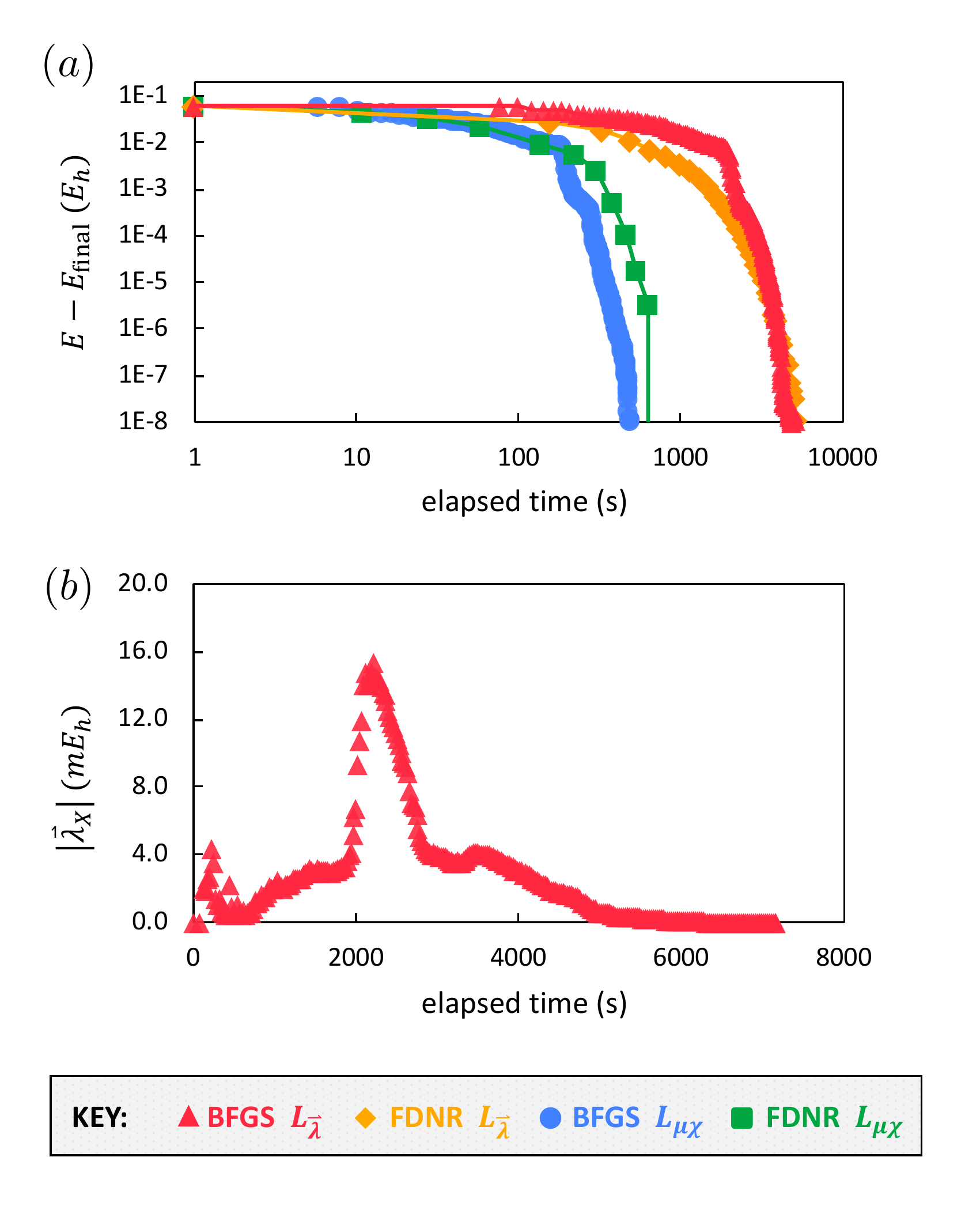}
    \caption{For the HOMO-LUMO singlet excitation in cc-pVDZ diazomethane, the difference between the current and converged excited state energy in Hartree is compared across optimization 
    strategies in (a) and the norm (in millihartrees) of the Lagrange multipliers associated with the orbital rotation parameters is shown over the course of an BFGS-$L_{\vec{\lambda}}$ 
    optimization in (b).}  
    \label{fig:timings_diazomethane}
\end{figure}

\begin{figure}
    \centering
    \includegraphics[scale=0.45]{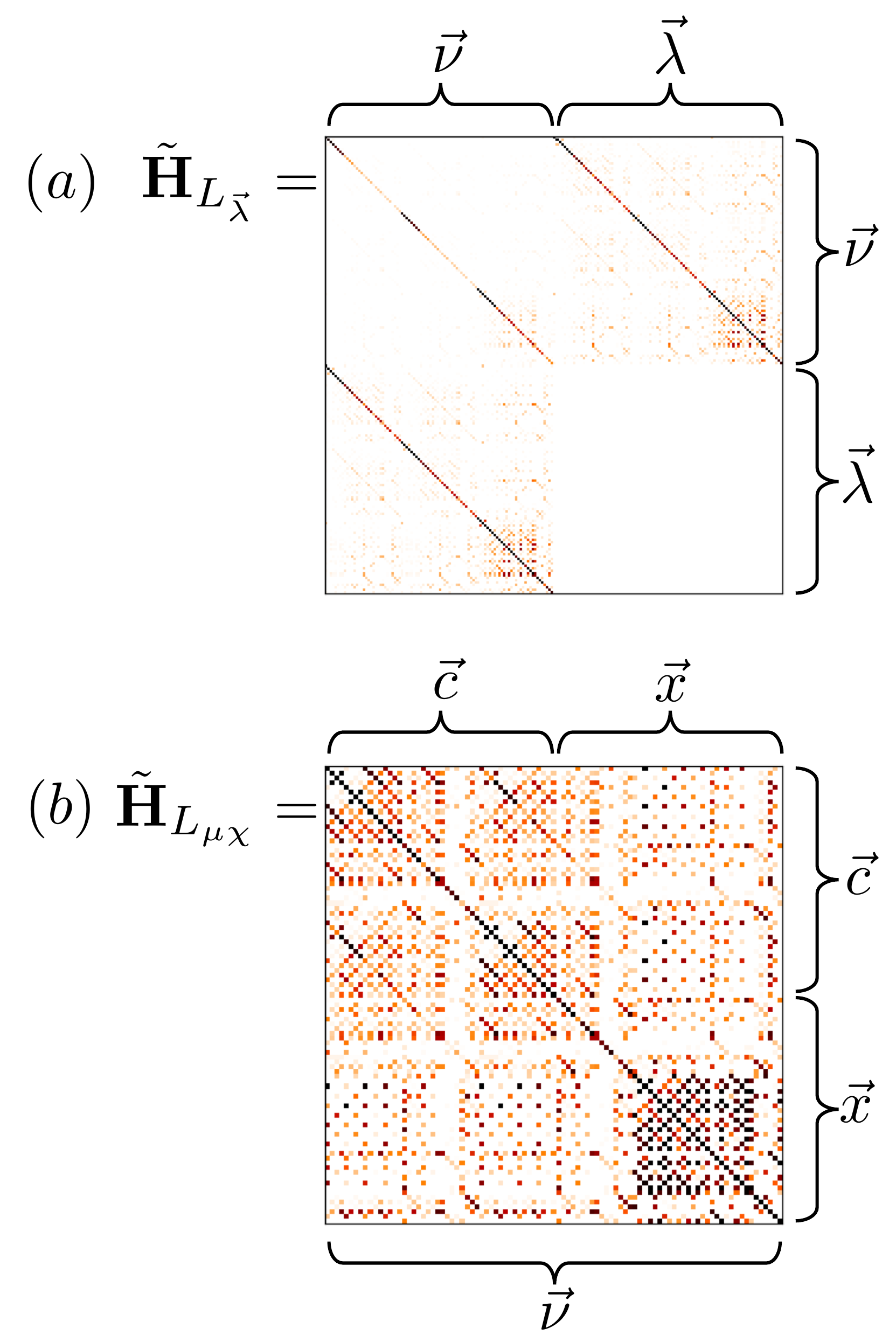}
    \caption{Heatmaps of the initial Hessian matrix for (a) the $L_{\vec{\lambda}}$ target function and (b) the $L_{\mu\chi}$ target function where $\mu=\frac{1}{2}$ and $\chi=1$ for the HOMO-LUMO 
    excitation in formaldehyde in the STO-3G basis. The values of the matrix are scaled such that the matrix elements equal to zero are white and the elements darken as they increase in magnitude. 
    Note that in order to emphasize detail, the Hessians are scaled according to $\mathbf{\tilde{H}_{ij}} = ( 1 -  \exp[-\big|\mathbf{H_{ij}}\big|] )$ and that $\mathbf{\tilde{H}}_{L_{\mu\chi}}$ is 
    enlarged with respect to $\mathbf{\tilde{H}}_{L_{\vec{\lambda}}}$. }
    \label{fig:hessian_heatmap}
\end{figure}

We now turn to the question of which strategy is most efficient when
optimizing the ESMF wave function.
Our key findings here are that the GVP-based
$L_{\mu\chi}$ objective function leads to faster
optimization than does $L_{\vec{\lambda}}$
and that, once using $L_{\mu\chi}$, the
efficiencies of the BFGS and FDNR methods become
system-dependent but similar.
Figures \ref{fig:timings_water}-\ref{fig:timings_diazomethane} show
four examples of this trend, which we have observed across all of the
systems we have tested.
In each of these four examples, roughly one order of magnitude
in speed is gained by moving to the $L_{\mu\chi}$ objective function.

To understand why the $L_{\vec{\lambda}}$ objective
function is less efficient, it is useful to analyze the
Lagrange multipliers.
At convergence, the energy will be stationary,
which in turn implies that the Lagrange multiplier
values will all be zero.
\begin{align}
     \frac{\partial L_{\vec{\lambda}}}{\partial \vec{\nu}} 
     = \frac{\partial}{\partial \vec{\nu}} \Big( W + \vec{\lambda}\cdot \frac{\partial E}{\partial \vec{\nu}} \Big) 
     &= \frac{\partial W}{\partial \vec{\nu}} + \mathbf{H}_E \vec{\lambda} = 0 
     \\
     \Rightarrow \hspace{2mm} \mathbf{H}_E \vec{\lambda} &=
     2(\omega-E)\frac{\partial E}{\partial \vec{\nu}} = 0
     \\
     \Rightarrow \hspace{2mm} \vec{\lambda} &= \vec{0}
\end{align}
Thus, we have guessed $\vec{\lambda}=0$ in our
optimizations.
However, as can be seen in the lower panels of
Figures \ref{fig:timings_water}-\ref{fig:timings_diazomethane},
the optimization moves the Lagrange multipliers
significantly away from zero before returning them there.
At the very least, this suggests that while our
initial guesses for the wave function parameters
and Lagrange multipliers are reasonable, they
are not particularly well paired, in that
the optimization of the wave function variables
drives the multipliers away from their optimal
values during a large fraction of the optimization.
This issue is simply not present when using
the $L_{\mu\chi}$ objective function as no
multipliers are present and thus no guess for them
is required.

To gain additional insight into the difficulties created
through the Lagrange multipliers, we can look at the
Hessian matrices produced by the two different objective
functions, for which examples are shown in 
Fig.\ \ref{fig:hessian_heatmap}.
As one might expect, the $L_{\vec{\lambda}}$ Hessian
has a blocked structure, with the multiplier-multiplier
block being zero trivially and the other three blocks being
diagonally dominant.
This structure is quite far from the identity-matrix guess
of standard BFGS, and although it may be possible to
construct a better estimate for the initial inverse Hessian
this would require evaluating at least some of the
second derivatives of the objective function individually,
which is not guaranteed to have the same cost scaling as
evaluating the energy.
Although good estimates may be achievable at low cost,
we have not in this study made any attempt at improving the
initial BFGS Hessian guess for either objective function,
and have simply used the identity matrix in both cases.
As Fig.\ \ref{fig:hessian_heatmap} shows, this very simple guess
is a better fit for the single-diagonal diagonally dominant
Hessian of the GVP-based objective function.
As we move towards a production-level implementation of the ESMF wave function and GVP, 
we hope to further improve the optimization algorithms. 
As the overall computational cost of ESMF is dominated by the number 
of Fock builds, we anticipate significant speedups through 
Hessian preconditioning, 
integral screening,\cite{Helgaker:2000_integral_screening} 
resolution of the identity approaches,\cite{Neese:2003:resolution_identity,Ren:2012:res_identity} and, as our objective function is invariant to some orbital rotations, 
geometric descent minimization methods.\cite{HF_Preconditioner:6:VanVoorhisGDM}

\begin{figure*}
\centering
\subfigure{
\resizebox{1 \textwidth}{!}{
\includegraphics{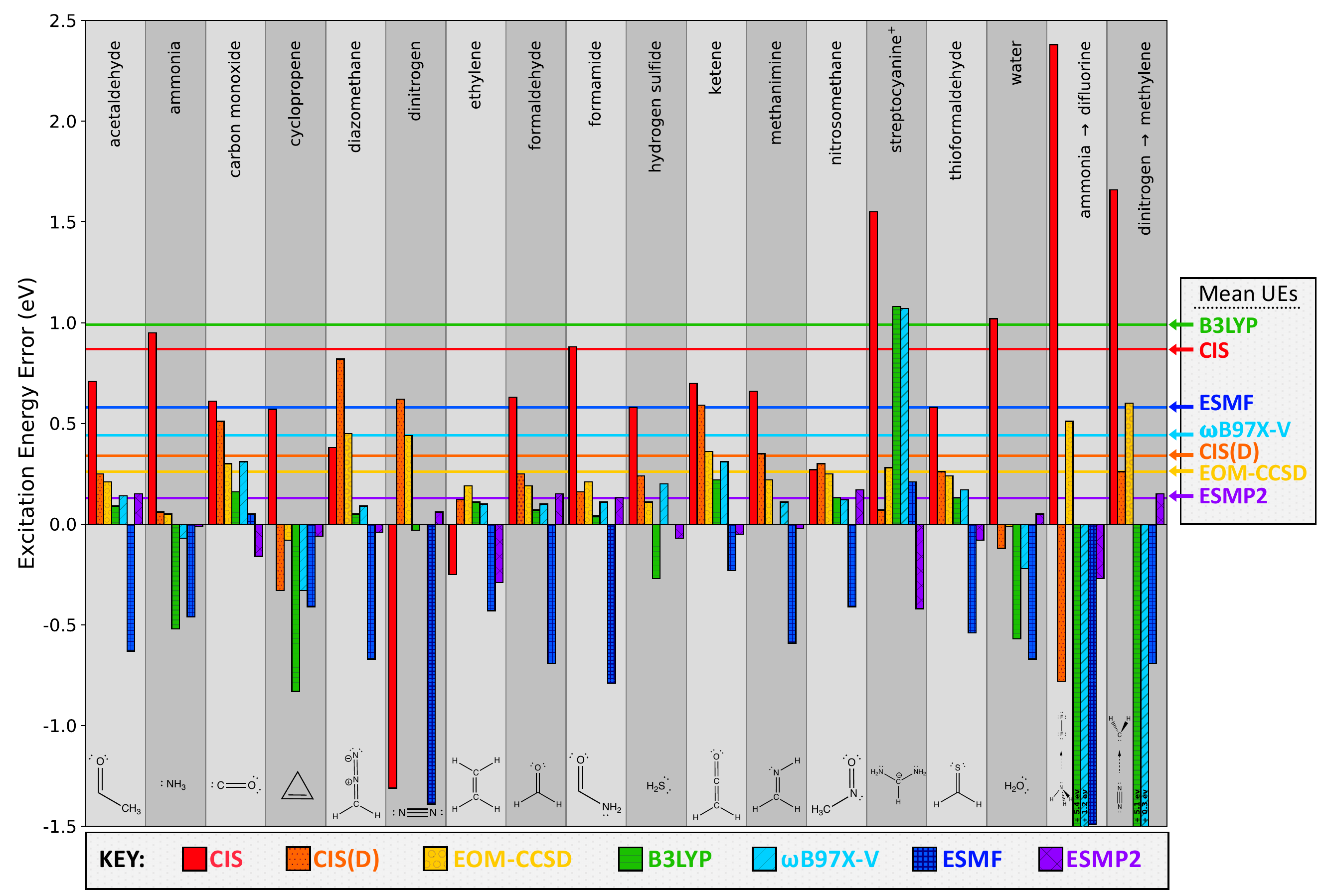}
}}
\newline
\subtable{
\resizebox{1 \textwidth}{!}{
\begin{tabular}{  L{0.1cm} R{1.0cm} L{1.0cm} R{0.8cm} @{.} L{0.8cm} R{0.8cm} @{.} L{0.8cm} R{0.8cm} @{.} L{0.8cm}  R{0.8cm} @{.} L{0.8cm} R{0.8cm} @{.} L{0.8cm} R{0.8cm} @{.} L{0.8cm} R{0.8cm} @{.} L{0.8cm}  }
 & \multicolumn{2} { l }{\textbf{Unsigned Errors} }          
 & \multicolumn{2} { c }{ \textbf{CIS}   }
 & \multicolumn{2} { c }{ \textbf{CIS(D)} }
 & \multicolumn{2} { c }{ \textbf{EOM-CCSD}   }
 & \multicolumn{2} { c }{ \textbf{B3LYP}   }
 & \multicolumn{2} { c }{ \textbf{$\omega$B97X-V}   }
 & \multicolumn{2} { c }{ \textbf{ESMF}  }
 & \multicolumn{2} { c }{ \textbf{ESMP2}  }  
 \\ \hline  \rule{0pt}{4ex}
 & \multicolumn{1}{l}{\textbf{Max}} 
 & \multicolumn{1}{l}{\textbf{(All)}}    &   2&38   &   0&82   &   0&60   &   6&91   &   2&69 &   1&49   &   0&42   \\
 & \multicolumn{1}{l}{\textbf{Mean}} 
 & \multicolumn{1}{l}{\textbf{(All)}}    &   0&87   &   0&34   &   0&26   &   0&99   &   0&44 &   0&58   &   0&13   \\
 \rule{0pt}{4ex}
 & \multicolumn{1}{l}{\textbf{Max}} 
 & \multicolumn{1}{l}{\textbf{(No CT)}}   &   1&55  &   0&82  &   0&45  &  1&08  &  1&07  &  1&39   & 0&42 \\
 & \multicolumn{1}{l}{\textbf{Mean}} 
 & \multicolumn{1}{l}{\textbf{(No CT)}}  &   0&73  &   0&32  &   0&22  &  0&27  &  0&22  &  0&51   & 0&12 
\end{tabular}
}}
\caption{ 
Singlet excitation energy errors from seven methods vs $\delta$-CR-EOM-CC(2,3)D in eV for
several small organic molecules.
From left to right for each molecule, the excitation energy error 
bars correspond to 
CIS (red/solid),
CIS(D) (orange/dots), 
EOM-CCSD (yellow/open circles), 
TDDFT/B3LYP (green/horizontal lines), 
TDDFT/$\omega$B97X-V (light blue/diagonal lines), 
ESMF (dark blue/gridlines), 
and ESMP2 (purple/diagonal gridlines).
For each method, 
we tabulate the maximum and mean unsigned 
errors across all 18 systems in the test set, 
\textbf{Max (All)} and \textbf{Mean (All)}, 
and the maximum and mean unsigned errors across 
only the 16 intramolecular excitations in the test set,
\textbf{Max (No CT)} and \textbf{Mean (No CT)}. 
Additionally, we plot each
\textbf{Mean (All)} UE with horizontal, solid lines 
in the plot.
For easier visualization of the data, we truncate the y-axis at -1.5 eV. As  the B3LYP and $\omega$B97X-V charge transfer excitation energy errors exceed this bound, 
we have included the absolute value of the error truncated by the axis.
See the SI for more information about the individual excitations,
but note that the two on the right are the two charge transfer cases.
}
\label{fig:excitation_energies}
\end{figure*}

\subsection{Benchmarking Excitation Energies}
\label{ssec::rsub_exen}

\begin{table}[]
    \centering
    \caption{
      Formal cost-scaling for methods used in this work.
      These scalings are with respect to the system size $N$ and
      are for canonical versions of the methods,
      i.e.\ without accelerations from two-electron integral screening
      or factorization.
    }
    \label{tb:scaling}
    \begin{tabular}{L{0.5cm} L{4cm} C{2.5cm} } 
        \hline \hline   \\
         & Method & Formal Scaling   \\
         \hline \rule{0pt}{4ex}
         & RHF\cite{Szabo-Ostland} & $N^{4}$  \\
         & CIS\cite{HeadGordon:2005:tddft_cis} & $N^{4}$ \\
         & TDDFT\cite{HeadGordon:2005:tddft_cis} & $N^{4}$  \\
         & ESMF  & $N^4$  \\
         & CIS(D)\cite{cisd:1994} & $N^{5}$  \\
         & EOM-CCSD\cite{piecuch:delta_eom_ccsd} & $N^6$  \\
         & $\delta$-CR-EOM-CC(2,3),D\cite{piecuch:delta_eom_ccsd} & $N^7$  \\
         & ESMP2 & $N^7$  \\
         \hline \hline
    \end{tabular}
\end{table}

We compiled a modest test set of small organic molecules that allows us to compare the accuracy of ESMF and ESMP2 against that of a range of single reference, weakly correlated excited state 
wave function methods. These methods include CIS, CIS(D), and notably EOM-CCSD and $\delta$-CR-EOM-CC(2,3),D, the latter of which scales as $\mathcal{O}(N^7)$ and is used as a 
high level benchmark.\cite{piecuch:delta_eom_ccsd} 
To contextualize the accuracy of ESMF and ESMP2 theories within the wider realm of excited state methods, we also present TDDFT benchmarks against $\delta$-CR-EOM-CC(2,3),D for 
both the B3LYP functional and the $\omega$B97X-V functional -- two popular hybrid GGA functionals.\cite{b3lyp:1,b3lyp:2,wB97X-V:initial}
For reference, the formal scaling of all methods used here is summarized in Table \ref{tb:scaling}.
While these scalings can in some cases be reduced
via sparse linear algebra or integral screening,\cite{HF:reduced_scaling,reduced_scaling:CIS_TDDFT,Goedecker:1999:linear_elec_struc}
we compare to canonical scalings here as our ESMF implementation does not yet
take advantage of such approaches.

Our test set includes a number of intramolecular
HOMO-LUMO singlet excitations as well as two long-range charge transfer
excitations,
NH\textsubscript{3}$(n)$ $\to$ F\textsubscript{2}$(\sigma^*)$
with a 6 \AA\ separation, and
N\textsubscript{2}$(\pi)$ $\to$ CH\textsubscript{2}$(2p)$
with a 10.4 \AA\ separation.
These types of CT excitations are known to cause difficulties for linear response theories; for example, CIS
fails to capture how the shapes and sizes of the
donor's and acceptor's orbitals change following the excitation.\cite{subotnik2011cis_ct}
In contrast, EOM-CCSD (through its doubles response operator)
and ESMP2 (through ESMF's variational optimization) do capture the
relaxation effects, which helps them achieve significantly better
(although not perfect) energetics in these CT cases.\cite{krylov:eomccsd,eomccsd:benchmark:geertson_bartlett:1989,eomccsd:benchmark:comeau_bartlett:1993,eomccsd:benchmark:stanton_bartlett:1993,eomccsd:benchmark:watts_bartlett:1996,Neuscamman:2016:var,Chris:2016:eom_hsvmc} 
Although the analysis for TDDFT is less straightforward,
even modern range-separated functionals do not account properly for
all orbital relaxation effects, which continue to produce difficulties
in charge transfer excitations\cite{chris:2019:vesdft,HeadGordon:2005:tddft_cis,tddft:charge_transfer_SIE} 
despite the clear improvements\cite{tddft:charge_transfer_nonlocal,wB97X-V:long_range_charge_transfer,tddft:mhg_30years,dft:long_range_fxls}
that range-separation offers.

The results for this survey are shown in 
Fig.\ \ref{fig:excitation_energies}
and tabulated in the SI.
Overall, we see that EOM-CCSD and ESMP2 
are most accurate in this test set, which we attribute
to their ability to provide fully excited-state-specific orbital
relaxations.
In contrast, CIS, which lacks proper orbital relaxation, has the
largest mean unsigned error (MUE) and maximum error of the wave function methods,
performing especially poorly in the two charge transfer systems.
Note that, although CIS can shape the orbitals for the electron and
hole involved in the excitation via superpositions of different
singles, it leaves the remaining occupied orbitals unrelaxed,
which is notably inappropriate in long-range CT where the large
changes in local electron densities should lead all nearby
valence orbitals to relax significantly.

CIS(D), an excited state analog of ground state MP2, recovers much of the electron correlation effects missing in CIS, halving the CIS MUE. 
Employing perturbation theory in the space of double excitations and using products of CIS single excitation amplitudes and ground-state MP2 
double excitation amplitudes to account for triple excitations from the ground state wave function, CIS(D) captures weak correlation and can 
improve the CIS excitation energies for only $\mathcal{O}(N^5)$ cost.\cite{cisd:1994}
The effect of the perturbative doubles and approximated triples on the excitation energies is certainly noticeable as the accuracy improves; 
however, CIS(D)'s use of ground state MP2 amplitudes leaves it lacking as from first principles, the electron-electron correlation in the 
ground state should differ from that in the excited state for any electrons involved in or near the excitation. We would thus expect a 
fully excited-state-specific perturbation theory to outperform CIS(D), and
indeed this is what we find.

Within this test set of systems, we unsurprisingly observed that the accuracy of TDDFT is both system and functional dependent. 
As summarized in Fig.\ \ref{fig:excitation_energies}, the accuracy of the B3LYP and $\omega$B97X-V functionals across the 
intramolecular valence-excitations rivals that of EOM-CCSD. 
Indeed, it is difficult to motivate moving away from the remarkably low computational cost TDDFT methods for such excitations.\cite{tddft:mhg_30years} 
However, the TDDFT results are not as accurate in the charge transfer systems. 
The B3LYP functional performs particularly poorly, and while the accuracy of TDDFT with the $\omega$B97X-V functional, 
which uses HF exchange at long-ranges, is not quite so catastrophic, it drastically underestimates the excitation energy by multiple eV.   

In this survey, ESMF consistently underestimates the excitation energy, and when the unsigned errors are compared, was more accurate than CIS, yet not as accurate as CIS(D).
The underestimation can be understood by recognizing that, in the excited state, ESMF captures the pair-correlation energy between the two electrons in open-shell orbitals, 
whereas no correlation at all
is preset in the RHF ground state (apart from Pauli correlation that of
course ESMF also has).
In addition, ESMF does in fact recover some weak correlation between different configurations of singly-excited determinants, 
i.e. for some configurations of $i \text{, } j\text{, } a \text{, and } b$, $\langle \Phi_i^a | e^{-\hat{X}} \hat{H} e^{\hat{X}} | \Phi_j^b \rangle \neq 0$. 
We are thus not surprised that ESMF excitation energies tend to be
underestimates due to this capture of some correlation. 
Note that, as this is a very incomplete accounting of correlation
effects (doubles and triples are missing), it is also not surprising
that the overall accuracy of ESMF is inferior to that of CIS(D), which
provides at least an approximate estimate of what the second order
correction for the doubles and triples should be.
Another notable
point about the accuracy of ESMF is that it is not
significantly different in the CT systems as compared to the
other systems, suggesting that it has successfully captured the
larger orbital relaxations present in CT.

Although its stand-alone accuracy leaves something to be desired, 
the ESMF wave function does provide an excellent starting point for
post-mean-field correlation theories, as evidenced by the
excellent performance of ESMP2.
Thanks to its orbital-relaxed starting point and excited-state-specific
determination of the doubles and triples, ESMP2 delivers the
highest overall accuracy when compared to the
$\delta$-CR-EOM(2,3)D benchmark.
In both intra- and inter-molecular excitations, ESMP2 is significantly more accurate than CIS, CIS(D), or ESMF, and
slightly more accurate than EOM-CCSD.
Of particular interest to note is that ESMP2 maintains its accuracy across both intramolecular valence excitations and long-range charge transfer excitations, and
while this test set is too limited and the basis set too small
to make strong recommendations,
this data suggests that ESMP2 may in some circumstances be
preferable to EOM-CCSD as well as TDDFT in both intra- and inter-molecular excitations.
Certainly the data motivate work on versions of ESMP2 that avoid using
the full set of uncontracted triples in the first order interaction space,
which should lower its cost-scaling.

\subsection{A Single-CSF Ansatz}
\label{ss:single_csf}

As many excitations are dominated by a single open-shell CSF, one might wonder
whether in these cases the full CIS-like CI expansion within ESMF is strictly necessary.
Although the effect of the remaining singly-excited CSFs is not negligible,
one could argue that their small weights put them firmly in the category of weak
correlation effects that should be handled by the perturbation theory.
For now, we have chosen not to pursue this direction in ESMP2 for two reasons.
First, it would limit the theory to single-CSF-dominated excitations.
Second, many of the other single excitations are much
closer in energy to the reference wave function than the doubles excitations are, thus
significantly increasing the risk of encountering intruder states.
That said, we have used our present implementation to test how much the absence of
these terms in ESMP2 matters if we restrict the reference function to
be a single CSF with optimized orbitals, which we will refer to here as the
oo-CSF ansatz.
\begin{align}
\label{eqn:oo-CSF}
    |\text{oo-CSF}\rangle = e^{\hat{X}} \Bigg( 
\hat{a}_{a\uparrow}^\dagger \hat{a}_{i\uparrow}|\Phi\rangle 
+ \eta \, 
\hat{a}_{a\downarrow}^\dagger \hat{a}_{i\downarrow}|\Phi\rangle  \Bigg).
\end{align}
As in Eq.\ (\ref{eqn:wfn}), $|\Phi\rangle$ denotes the RHF solution.
However, the definition of $\hat{X}$ is slightly different.
The oo-CSF ansatz is invariant to occupied-occupied and virtual-virtual orbital
rotations that do not involve orbitals $i$ or $a$, but such rotations that do
involve these orbitals now matter, and so we have enabled these portions of the
$X$ matrix in addition to the ESMF occupied-virtual block
shown in Fig.\ \ref{fig:X_matrix}(b).
Finally, note that $\eta$ is not a variable and is simply set to $1$ if we
wish to work with the spin singlet and $-1$ for the triplet.

As seen in Table \ref{tb:csf_vs_full_cis}, we tested oo-CSF as a reference
for ESMP2 in three systems where the structure of the optimized ESMF wave function
suggested that oo-CSF had a good chance of being effective and one in
which it did not appear appropriate.
For water, the ESMF wave function is already dominated by a single CSF.
For formaldehyde and methanimine, the additional subset of
occupied-occupied rotations that we enabled for oo-CSF allow the primary
components of the excitation to be converted into a single CSF by
mixing the ESMF HOMO with the other occupied orbitals.
While this simplification is certainly not always possible
(N$_2$ is a good counterexample, having two large components involving
completely separate sets of molecular orbitals)
our results suggest that when it is, the absence of the other singles
excitations in our ESMP2 method may not be of much consequence.
In the future, the efficacy of oo-CSF for single-CSF-dominated states could
perhaps be exploited in a couple of different ways.
On the one hand, it is a simpler ansatz and so may prove easier to optimize than
ESMF, which even in systems where secondary CSFs were not negligible could be useful
if it provides a low-cost, high-quality initial guess for the ESMF optimization.
On the other hand, its simpler structure could prove useful in simplifying
the implementation of ESMP2.

\begin{table}[t]
\caption{
  Excitation energy errors in eV relative to $\delta$-CR-EOM-CC(2,3),D 
  for the HOMO$\rightarrow$LUMO singlet excitations
  of water, formaldehyde, methanimine, and dinitrogen.
  Below the name of each molecule, we report the CSF coefficients in the ESMF wave
  function with amplitudes larger than 0.1. 
}
\label{tb:csf_vs_full_cis}
\begin{tabular}{ L{0.01cm} R{2.0cm} @{.} L{2.0cm} R{0.2cm} @{\textsubscript{.}} L{0.8cm}  R{0.4cm} @{.} L{0.4cm} R{0.4cm} @{.} L{0.4cm} R{0.6cm} @{.} L{0.6cm} R{0.6cm} @{.} L{0.6cm} }
   \hline \hline \\
  & \multicolumn{2}{c}{ }
  & \multicolumn{2}{c}{ }
  & \multicolumn{2}{c}{ }
  & \multicolumn{2}{c}{oo-}
  & \multicolumn{2}{c}{ESMP2}
  & \multicolumn{2}{c}{ESMP2} 
  \\
  &\multicolumn{2}{l}{  }
  &\multicolumn{2}{l}{  }
  &\multicolumn{2}{c}{ESMF}
  &\multicolumn{2}{c}{CSF}
  &\multicolumn{2}{c}{w/ESMF}
  &\multicolumn{2}{c}{w/oo-CSF}
  \\
  \hline \\
  \rule{0pt}{4ex}
  & \multicolumn{2}{l}{Water}     &\multicolumn{2}{l}{  } & -0&67  & -0&66  &  0&05  & 0&06 \\
  & \multicolumn{2}{l}{  \textsubscript{HOMO} ${}_{\rightarrow}$ \textsubscript{LUMO} }
  & \textsubscript{0} & \textsubscript{70}
  & \multicolumn{8}{c}{ }
  \\
  \rule{0pt}{4ex}
  & \multicolumn{2}{l}{Formaldehyde}     &\multicolumn{2}{l}{  } & -0&69  & -0&66  &  0&15  & 0&18 \\
  %& \multicolumn{2}{l}{  \textsubscript{HOMO   }${}_{\,\,\,\rightarrow}$ \textsubscript{LUMO} }
  & \multicolumn{2}{l}{  \textsubscript{HOMO} ${}_{\rightarrow}$ \textsubscript{LUMO} }
  & \textsubscript{0} & \textsubscript{66}
  & \multicolumn{8}{c}{ }
  \\
  %& \multicolumn{2}{l}{  \textsubscript{HOMO-3}${}_{\,\rightarrow}$ \textsubscript{LUMO} }
  & \multicolumn{2}{l}{  \textsubscript{HOMO-3} ${}_{\rightarrow}$ \textsubscript{LUMO} }
  & \textsubscript{0} & \textsubscript{22}
  & \multicolumn{8}{c}{ }
  \\
  \rule{0pt}{4ex}
  & \multicolumn{2}{l}{Methanimine}    &\multicolumn{2}{l}{  }  & -0&59  & -0&55  &  -0&02  & 0&02 \\
  %& \multicolumn{2}{l}{  \textsubscript{HOMO  }${}_{\,\rightarrow}$ \textsubscript{LUMO} }
  & \multicolumn{2}{l}{  \textsubscript{HOMO} ${}_{\rightarrow}$ \textsubscript{LUMO} }
  & \textsubscript{0} & \textsubscript{67}
  & \multicolumn{8}{c}{ }
  \\
  & \multicolumn{2}{l}{  \textsubscript{HOMO-2} ${}_{\rightarrow}$ \textsubscript{LUMO} }
  & \textsubscript{-0} & \textsubscript{17}
  & \multicolumn{8}{c}{ }
  \\
  \rule{0pt}{4ex}
  & \multicolumn{2}{l}{Dinitrogen}    &\multicolumn{2}{l}{  }  & -1&39  & -1&13 &  0&06 & 0&52 \\
  %& \multicolumn{2}{l}{  \textsubscript{HOMO-1  }${}_{\,\rightarrow}$ \textsubscript{LUMO} }
  & \multicolumn{2}{l}{  \textsubscript{HOMO-1} ${}_{\rightarrow}$ \textsubscript{LUMO} }
  & \textsubscript{0} & \textsubscript{49}
  & \multicolumn{8}{c}{ }
  \\
  & \multicolumn{2}{l}{  \textsubscript{HOMO} ${}_{\rightarrow}$ \textsubscript{LUMO+1} }
  & \textsubscript{0} & \textsubscript{49}
  & \multicolumn{8}{c}{ }
  \\
 \rule{0pt}{1ex} \\ \hline \hline
 \end{tabular}
\end{table}

\subsection{Targeting with Other Properties}
\label{ss:lih_property_vec}

\begin{figure}
\centering
\includegraphics[scale=0.34]{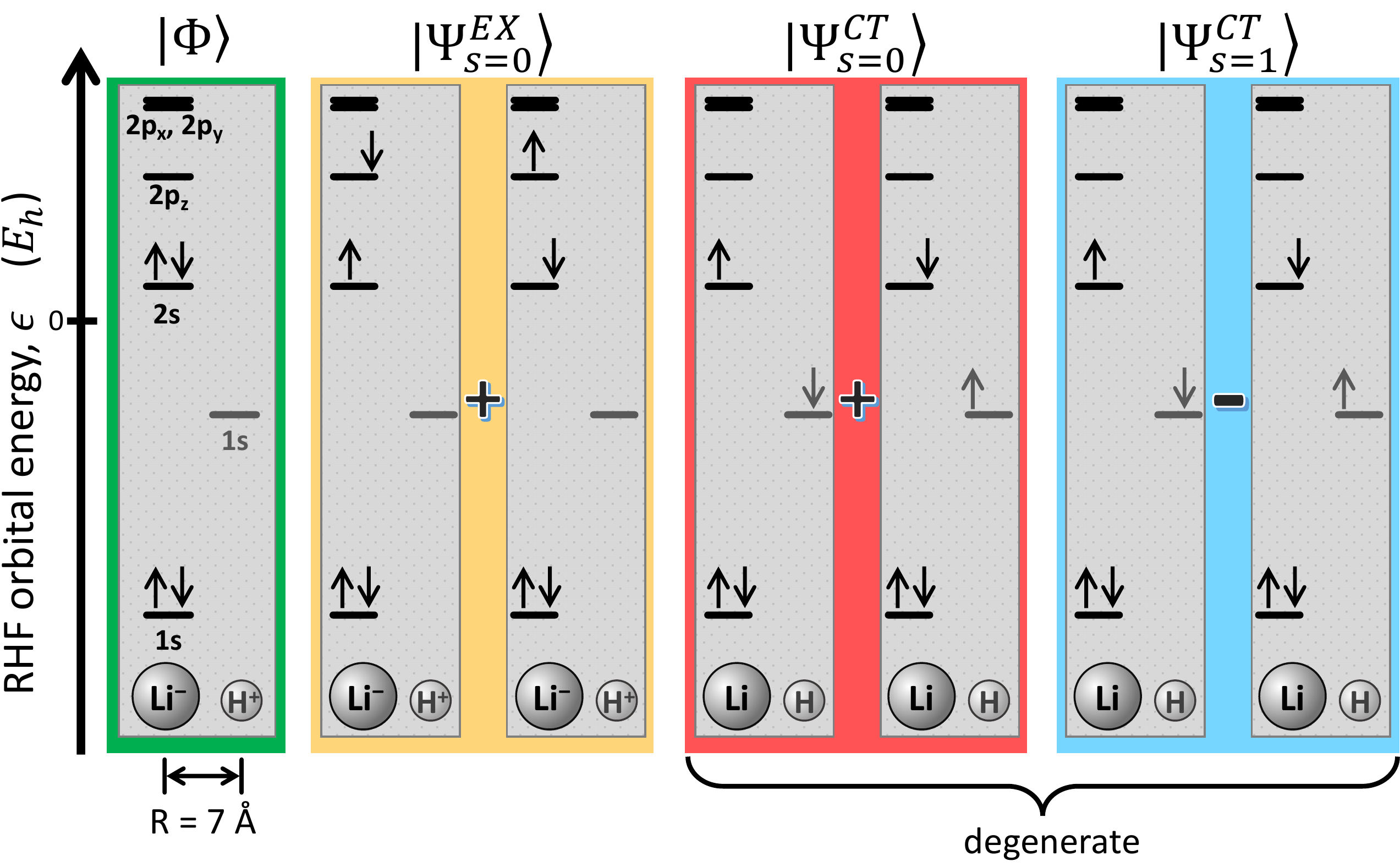}
\caption{
  Molecular orbital diagrams for the key determinants in the
  four relevant states in our LiH test.
  From left to right, we have the ionic RHF wave function $|\Phi\rangle$, and
  the main CSFs that contribute to the singlet state $|\Psi^{EX}_{s=0}\rangle$ that maintains the ionic
  character,
  and that contribute to the degenerate singlet and triplet states
  $|\Psi^{CT}_{s=0}\rangle$ and $|\Psi^{CT}_{s=1}\rangle$ in which
  neutrality has been restored by an Li$\rightarrow$H charge transfer.
  Note that while the molecular orbitals are arranged based on their RHF orbital energies, the energy gaps are not to scale.
}
\label{fig:LiH_states}
\end{figure}

So far, we have focused on how an energy-targeting GVP can improve
ESMF optimizations.
We now turn our attention to the use of other properties to improve
the robustness of optimization in the face of poor initial guesses,
energetic degeneracy, and poor energy targeting.
To investigate these aspects of the GVP, we study
stretched LiH (bond distance 7 \AA) in the STO-3G basis,
whose low-lying states can be seen in Fig.\ \ref{fig:LiH_states}.
The idea is to optimize to the $|\Psi^{CT}_{s=0}\rangle$ state
despite the challenges of
(a) initial guesses that contain varying mixtures of
$|\Psi^{EX}_{s=0}\rangle$ and $|\Psi^{CT}_{s=0}\rangle$ character,
(b) setting the energy targeting to aim at the wrong state,
namely setting $\omega$ to the ESMF energy for 
$|\Psi^{EX}_{s=0}\rangle$, and
(c) the presence of $|\Psi^{CT}_{s=1}\rangle$, which is
energetically degenerate with $|\Psi^{CT}_{s=0}\rangle$ at
this bond distance.
While the latter difficulty could be resolved by constraining our
CI coefficients to produce only singlet states, we intentionally
leave our CI coefficients unconstrained.
Instead, we will investigate the efficacy of overcoming the challenges
of degeneracy, poor $\omega$ choice, and poor initial guesses by
including additional properties in the GVP's deviation vector $\vec{d}$.

\begin{figure}
\centering
\includegraphics[scale=0.45]{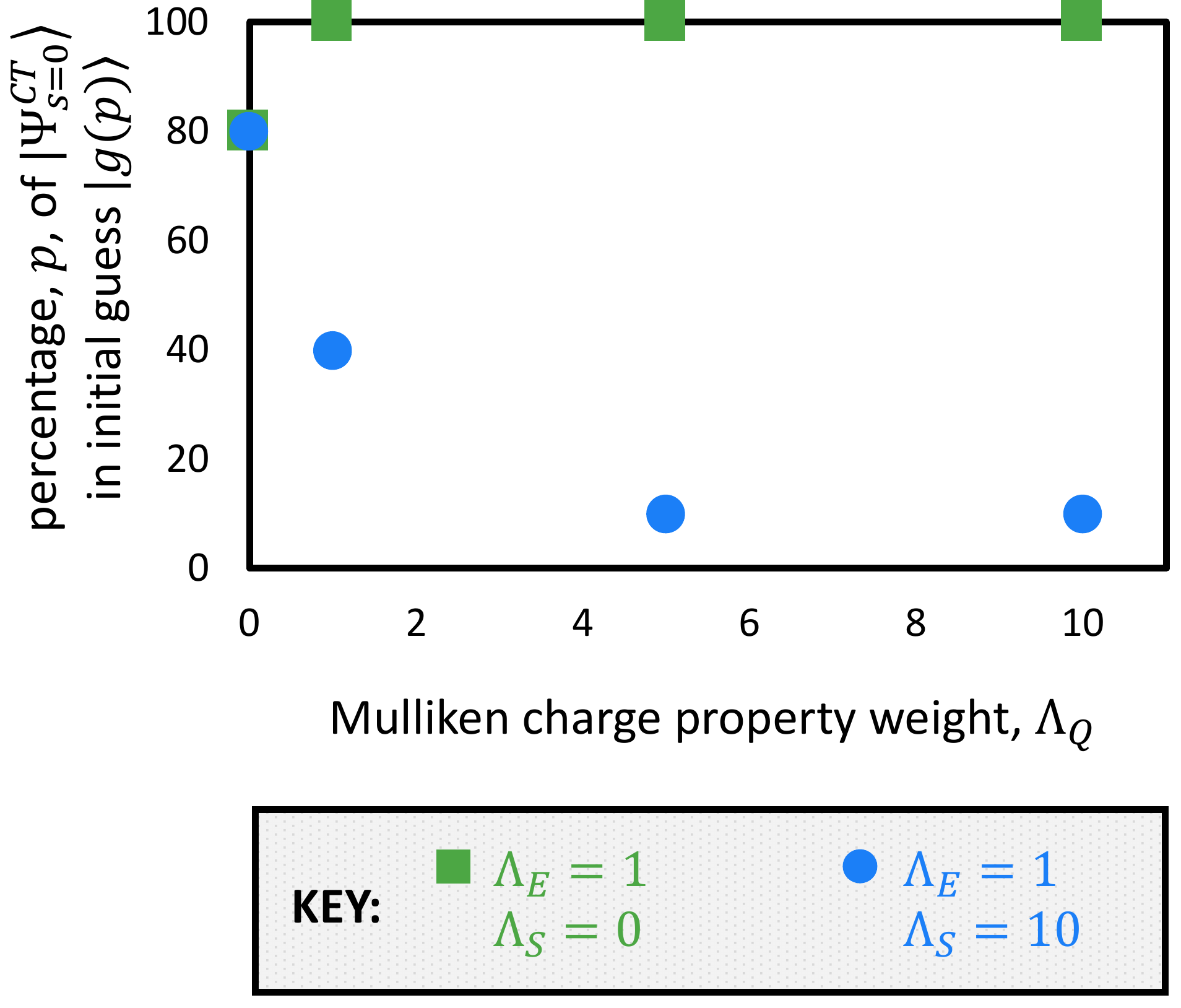}
\caption{
  For various choices of the $\Lambda$ matrix weights, we report the minimum
  value of $p$ required for our initial guess $|\mathrm{g}(p)\rangle$ to
  successfully optimize to the correct
  energy for the targeted $|\Psi^{CT}_{s=0}\rangle$ state.
}
\label{fig:mc_wt_plot}
\end{figure}

As we wish to arrive at the neutral $|\Psi^{CT}_{s=0}\rangle$ state while
avoiding the corresponding triplet state and being resilient to
an initial guess contaminated by the ionic $|\Psi^{EX}_{s=0}\rangle$ state,
the total spin and the Mulliken charges\cite{Szabo-Ostland}
of the atoms are obvious
candidates for additional properties that should help uniquely identify
our target state.
We therefore chose our property deviation vector as
\begin{align}
  \vec{d} = \left\{ \hspace{1mm}
  \left< E \right> - \omega, \hspace{1mm}
  \left< Q_{Li} \right> - \eta, \hspace{1mm}
  \sqrt{\left< S^2 \right>} - \zeta  \hspace{1mm}
   \right\}
\end{align}
where $\left<Q_{Li}\right>$ is the Mulliken charge on the Li atom
and we set $\eta=0$ and $\zeta=0$ so as to target a neutral singlet.
Happily, both the values and the derivatives of the
Mulliken charges and the total spin
\begin{align}
    \langle S^2 \rangle =
    \frac{\sum_{ia} (\sigma_{ia} - \tau_{ia})^2}
         {\sum_{ia} \sigma_{ia}^2 + \tau_{ia}^2}
\end{align}
are easily evaluated for a CIS-like wave function like ESMF, and so the
use of these properties does not change the cost-scaling of the method.
In order to conveniently study the effects of putting different amounts
of emphasis on different property deviations, we modify the GVP objective
function to take the following form.
\begin{align}
  \label{eqn:weighted_objective}
  L_{\Lambda} =
  \mu \hspace{1.5mm} \vec{d}^{\,\,T} \Lambda \hspace{1.5mm} \vec{d}
  + (1-\mu) |\nabla E|^2 \\
\label{eqn:lambda_weight_matrix}
\Lambda = \begin{bmatrix}
\Lambda_E & 0 & 0 \\
0 & \Lambda_Q &  0 \\
0 & 0 & \Lambda_S 
\end{bmatrix}
\hspace{10mm}
\end{align}
Of course, this is equivalent to setting the semi-positive-definite
matrix $\Lambda$ to unity and
scaling the definitions of the different properties, but we
find the above form more convenient for presenting the different
relative weightings that we placed on our three different
property deviations.

For our initial guess $|\mathrm{g}(p)\rangle$,
we have set the orbital basis to the the RHF orbitals
and have used varying mixtures of
$|\phi^{EX}\rangle$ and $|\phi^{CT}\rangle$,
which are the CIS wave functions corresponding to  
$|\Psi^{EX}_{s=0}\rangle$ and $|\Psi^{CT}_{s=0}\rangle$,
respectively.
\begin{align}
    \label{eqn:guess_wave_function}
    |\mathrm{g}(p)\rangle =
    \sqrt{\frac{p}{100}} \;\, \Big| \phi^{CT}\Big\rangle
  + \sqrt{\frac{100-p}{100}} \;\, \Big|\phi^{EX}\Big\rangle
\end{align}
For each choice of the property weights in the $\Lambda$ matrix,
we tested whether $|\mathrm{g}(p)\rangle$ would successfully converge
to the $|\Psi^{CT}_{s=0}\rangle$ state for the cases $p$ = 
0, 10, 20, $\ldots$, 100.
Each optimization was performed via FDNR minimization of $L=\chi L_{\Lambda} + (1-\chi) E$, with $\mu$
stepped down from one to zero by intervals of 0.1 and $\chi$ switched from one to zero on the twentieth FDNR iteration.
A value of -7.146 $E_h$, the ESMF energy for $|\Psi^{EX}_{s=0}\rangle$, was used for $\omega$ throughout.

As seen in Fig.\ \ref{fig:mc_wt_plot}, placing significant weights
on both the charge and spin deviations allows for successful optimizations
even with very poor initial guesses and our intentionally off-center value
for $\omega$.
With $\Lambda_S=10$ and $\Lambda_Q > 4$, we find that having as little
as 10\% of the correct CIS wave function in the initial guess leads to a
successful optimization.
When we do not include the spin targeting (i.e.\ when we set
$\Lambda_S=0$), we find that the charge targeting is much
less effective, with no optimizations succeeding when less than 80\%
of the correct CIS wave function is in the guess, regardless of the value of $\Lambda_Q$.
This result was somewhat unexpected, given that our guess is a pure
spin singlet.
We had expected that by giving the optimization a strong preference for
neutral states, we would have converged to a linear combination of
$|\Psi^{CT}_{s=0}\rangle$ and $|\Psi^{CT}_{s=1}\rangle$ that, while
perhaps displaying some spin contamination, at least had the correct energy.
Instead, we find that using spin to break the optimization degeneracy
(by setting $\Lambda_S=10$) is essential for robust convergence.

\vspace{6mm}
\section{Conclusions}
\label{sec:conclusion}

We have presented a generalization of the variational principle based on
the energy gradient and the idea of constructing a flexible system
for optimizing a state that can be specified uniquely by a list of properties.
This approach is formally exact while avoiding the difficulties associated with 
squaring the Hamiltonian operator.
Instead, it demands that a limited amount of energy second derivative information
be evaluated, but, and this point is crucial, the required derivatives do not
lead to an increase in cost scaling compared to the traditional ground state
variational principle.
So long as the properties used to identify the desired state do not themselves
lead to an increase in cost scaling, the approach is therefore expected to
maintain the scaling of its ground state counterpart.

Combining these ideas with excited state mean field theory, we have shown
that the latter's optimization can be carried out without the need for
the Lagrange multipliers that were present in its original formulation.
We find that this approach leads to substantial
efficiencies in the optimization thanks to both a simpler Hessian and
an objective function that is bounded from below and thus easier to use
straightforwardly with quasi-Newton optimization methods.
We have also shown that a full Newton-Raphson approach can be realized
efficiently and without Hessian matrix construction by formulating
Hessian matrix-vector products approximately via a finite-difference
of gradients.
Although it is not yet clear whether quasi-Newton methods or this
full Newton approach will ultimately be faster for excited state
mean field optimizations, what is clear is that the objective function
based on the generalized variational principle is strongly preferable
to the original objective function that relied on Lagrange multipliers.

With the ability to converge excited state mean field calculations in
a larger set of molecules than was previously possible, we compared
the corresponding second order perturbation theory
to other commonly-used single-reference excited state methods and
found its accuracy to be highly competitive.
This success motivates both work on an internally contracted version
of this perturbation theory in order to reduce its cost scaling
and on fully excited-state-specific coupled cluster methods, which,
if the history of ground state investigations is any guide, should
be even more reliable than the perturbation theory.

More broadly, the generalized variational principle appears to offer
new opportunities in many different areas of electronic structure theory.
The ability to use a property vector to define which state is being
sought without changing the final converged wave function should be
especially useful in multi-reference investigations, where root flipping
often prevents excited-state-specific calculations.
By combining the energy with other properties,
we demonstrated that the GVP could be used to resolve
an individual state even in the presence of degeneracy,
poor initial guesses, and poor energy targeting.
There are of course many properties one could explore, 
but some that come immediately to mind are the dipole moment, changes in bond order from the ground state, 
and the degree of overlap with wave function
estimates from other methods such as state averaging.
In addition to multi-reference theory, the generalized variational
principle appears to offer a route to defining exact density functionals
for excited states so long as those states can be specified uniquely
by a list of properties.
Combined with promising preliminary data from a density functional
extension of excited state mean field theory,\cite{chris:2019:vesdft}
this formal foundation
may allow for interesting new directions in density functional development.

\section{Acknowledgements}

We thank Piotr Piecuch for helpful discussions.
This work was supported by the National Science Foundation's
CAREER program under Award Number 1848012.
Development, testing, and timing calculations were performed using
the Berkeley Research Computing Savio cluster.
The more extensive calculations required for the energy comparison between methods were
performed using the National Energy Research Scientific Computing Center, a DOE Office of Science User Facility 
supported by the Office of Science of the U.S. Department of Energy under Contract No. DE-AC02-05CH11231.

J.A.R.S. acknowledges that this material is based upon work supported by the National Science Foundation Graduate 
Research Fellowship Program under Grant No. DGE 1752814. Any opinions, findings, and conclusions or recommendations 
expressed in this material are those of the author(s) and do not necessarily reflect the views of the National Science Foundation.

\section{Supporting Information}
\label{sec:SI_paragraph}

The supporting information includes technical details about the FDNR numerical optimization, descriptions of the 
excitations in the test set, tabulated results for excitation energy errors, and molecular geometries for the 
systems tested in this work. This information is available free of charge via the Internet at http://pubs.acs.org.

\providecommand{\latin}[1]{#1}
\providecommand*\mcitethebibliography{\thebibliography}
\csname @ifundefined\endcsname{endmcitethebibliography}
  {\let\endmcitethebibliography\endthebibliography}{}

\end{document}